\documentclass[12pt]{article}
\usepackage{graphicx} % Required for inserting images
\usepackage{physics}
\usepackage[dvipsnames]{xcolor}
\usepackage{ragged2e}
\usepackage{hyperref}
\usepackage[in]{fullpage}
\usepackage{amssymb}
\usepackage{amsmath}
\usepackage{amsfonts}
\usepackage{mathrsfs}

\usepackage{authblk}

\newcommand{\AJS}[1]{} %\textcolor{Orange}{}{\textsf{[AJS: #1]}} }
\newcommand{\be}{\begin{equation}}
\newcommand{\bea}{\begin{eqnarray}}

\newcommand{\bc}[1]{}
\newcommand{\ec}[1]{}
\newcommand{\deltatau}{\tau_*}

\newcommand{\beq}{\begin{equation}}
\newcommand{\eeq}{\end{equation}}
\newcommand{\irs}{\ell_{\text{IR}}}
\numberwithin{equation}{section}

\DeclareMathOperator{\Shi}{Shi}

\title{Quantum Fluctuations of the Black Hole Horizon}

\author[1,2]{Ben Freivogel\thanks{freivogel@uva.nl}}
\author[1]{Antony J. Speranza\thanks{asperanz@gmail.com}}
\author[1]{Erik Verlinde\thanks{verlinde@uva.nl}}

\affil[1]{\small \it Institute for Theoretical Physics, University of Amsterdam, Science Park 904, 1098 XH, Amsterdam, The Netherlands}

\affil[2]{\small \it GRAPPA, University of Amsterdam, Science Park 904, 1098 XH, Amsterdam, The Netherlands}

\date{June 26, 2026}

\begin{document}

\maketitle

\begin{abstract}
Classical black holes have sharply defined event horizons,
but quantum mechanically the horizon acquires a quantum uncertainty, called the `quantum width' by Marolf. 
We propose a definition of the quantum width by a physical
experiment involving the last moment a signal emitted from an ingoing 
light ray can escape to infinity.  
% We define and calculate the quantum width of spherically symmetric black holes in perturbative quantum gravity. 
Calculations of this observable for spherically 
symmetric black holes in perturbative quantum 
gravity reveal that the quantum
width
% We find that the quantum width 
depends on the resolution of the probe, and is often much larger than the Planck scale. For example, for Schwarzschild black holes in four dimensions in a particular regime of parameters, a piece of the horizon of size $\sigma_\perp$ has quantum width roughly $\sqrt{l_P r_s^2/\sigma_\perp}$.
%We calculate quantum fluctuations of black hole horizons within perturbative quantum gravity. We give a gauge-invariant definition of the horizon fluctuations. The fluctuations depend on the resolution of the probe, and in many regimes are much larger than the Planck scale. 

\end{abstract}

\flushbottom

\newpage

% \title{Quantum Widths of the BH Horizon}
% \author{Ben Freivogel, Antony Speranza, and Erik Verlinde}
% \date{}

% \begin{document}

% \maketitle

\section{Introduction}

It is generally accepted that gravity is quantised and that as a  result the space-time geometry fluctuates. Computing the size of these spacetime fluctuations turns out to be surprisingly difficult. 
Perturbative quantum gravity naively suggests that fluctuations in the spacetime geometry only happen at the Planck scale, and hence are impossible to observe. 
This argument is partly based on dimensional analysis and on the assumption that the only divergences happen in the UV and are regularised by a Planckian cut off. 
This suggests that the size of the fluctuations in the spacetime geometry is determined by the only UV scale in the problem. This reasoning overlooks a number of 
issues that are relevant for the question at hand. 

First of all, physical observables in quantum gravity are generally non-local, since they need to be defined in a generally covariant way and as a result must obey constraints that impose diffeomorphism invariance. 
The second related issue is the appearance of infrared divergences. A physical setup that is designed to measure the size of fluctuations in the spacetime geometry generally involves an IR scale, and the relevant physical observable can in principle be dependent on the scale.
This also implies that if we send this IR scale to infinity that the relevant physical quantities could diverge. The size of the fluctuations can thus be determined by a combination of the UV and IR scales that occur in the problem.

In this paper we are interested in a particular observable of this kind, which we call “the quantum width of the horizon.”  This name appeared for the first time in an analysis of Marolf, who studied the uncertainty in the location of the black hole horizon due to local quantum fluctuations in the Schwarzschild geometry \cite{Marolf2003}.  Following physical arguments that involved  ideas from black hole thermodynamics, Marolf found that the quantum width of the horizon in four spacetime dimensions is equal to the geometric mean of the Planck scale and the size of the black hole horizon. 
In other words, in this case the size of the fluctuations in the spacetime geometry gives a result that is larger than the Planck scale and also depends on the relevant IR scale. Our goal is to revisit the study of the quantum width of the horizon by setting up a precise physical measurement and computing the fluctuations in the corresponding physical observable using perturbative quantum gravity techniques.

The Marolf computation estimates the effects from the ``thermal atmosphere'' of the black hole. These can be thought of as a one-loop correction to the background geometry. The loops may include matter fields, photons, gravitons, etc. However, we want to begin by computing the ``tree level'' effect, namely, the fluctuations in the horizon due to gravitational perturbations. This approach is similar to the work of Parikh and Pereira \cite{Parikh:2024zmu} where they calculate the fluctuations in the area of the horizon. In this work, we will give a precise definition of the location of the horizon in the presence of fluctuations, and we focus on the quantum width rather than the fluctuations in the area.

An important question that we would like to address is whether the size of the fluctuations can be understood using holographic or thermodynamic arguments. One of the important physical quantities in the problem is the Bekenstein-Hawking entropy associated to the horizon. Simple thermodynamic arguments suggest that in the canonical ensemble the mass $M$ of the black hole, and hence the Schwarzschild radius $r_s$, fluctuates by an amount 
\begin{equation}
{\Delta r_s\over r_s} \sim {\Delta M\over M}   \sim {1\over \sqrt{S}}.
\end{equation}
The quantum width $L$ of the horizon is not simply equal to the fluctuations in the Schwarschild radius. Instead, we define $L$ as the geodesic distance between $r_s$ and $r_s+\Delta r_s$. Using the near-horizon metric, the relation is
\begin{equation}
\label{thermo-estimate}
L^2 \sim  {\Delta r_s\over T} \sim {1\over \sqrt{S}}\, {r_s\over T}
\end{equation}
where $T$ denotes the Hawking temperature. 
%This leads to the estimate that the quantum width $L$ is controlled by the entropy $S$ according to 
%\begin{equation}
%    L^2 
%\end{equation}
Similar reasoning was followed by Zurek and Verlinde in their study of spacetime fluctuations in causal diamonds and agrees in four dimensions with Marolf's result. 

In higher dimensions, the two results do not agree, which is one motivation for revisiting Marolf's calculation. In addition, we expect that the fluctuations of the horizon should depend on the length and time scale at which it is probed. Finally, it is important to define the quantum width carefully in a gauge-invariant way.

\subsection{Setup and summary of results}

The physical set-up is as follows: we imagine sending in a massless probe from infinity towards the black hole along a null geodesic. At a certain affine time $U^*$, the probe sends a signal that is received at infinity.
Our goal is to answer the question: what is the uncertainty in the affine time $\lambda$ that it takes for the probe to reach the horizon? %We will normalize the affine parameter and fix the initial conditions of the geodesic at infinity, where gravitational fluctuations vanish.

In physical terms,  this means: what is the last affine time at which the probe is able to send a signal to the asymptotic observer?
The physical observable is thus expressed as an integral along the null geodesic in a geometry that is allowed to fluctuate. 
Concretely,  this means that we calculate the affine distance to the event horizon of the black hole in the presence of a metric perturbation.
We assume that gravitational perturbations fall off sufficiently fast near infinity that we can fix initial conditions for the null geodesic and the normalization of the affine parameter near infinity. We will check later whether this assumption is self-consistent.  After calculating the perturbation in the affine parameter $\Delta \lambda$, we use perturbative quantum gravity to estimate the variance $\expval{(\Delta \lambda)^2}$. 

\begin{figure}
    \centering
    \includegraphics[scale=0.5]{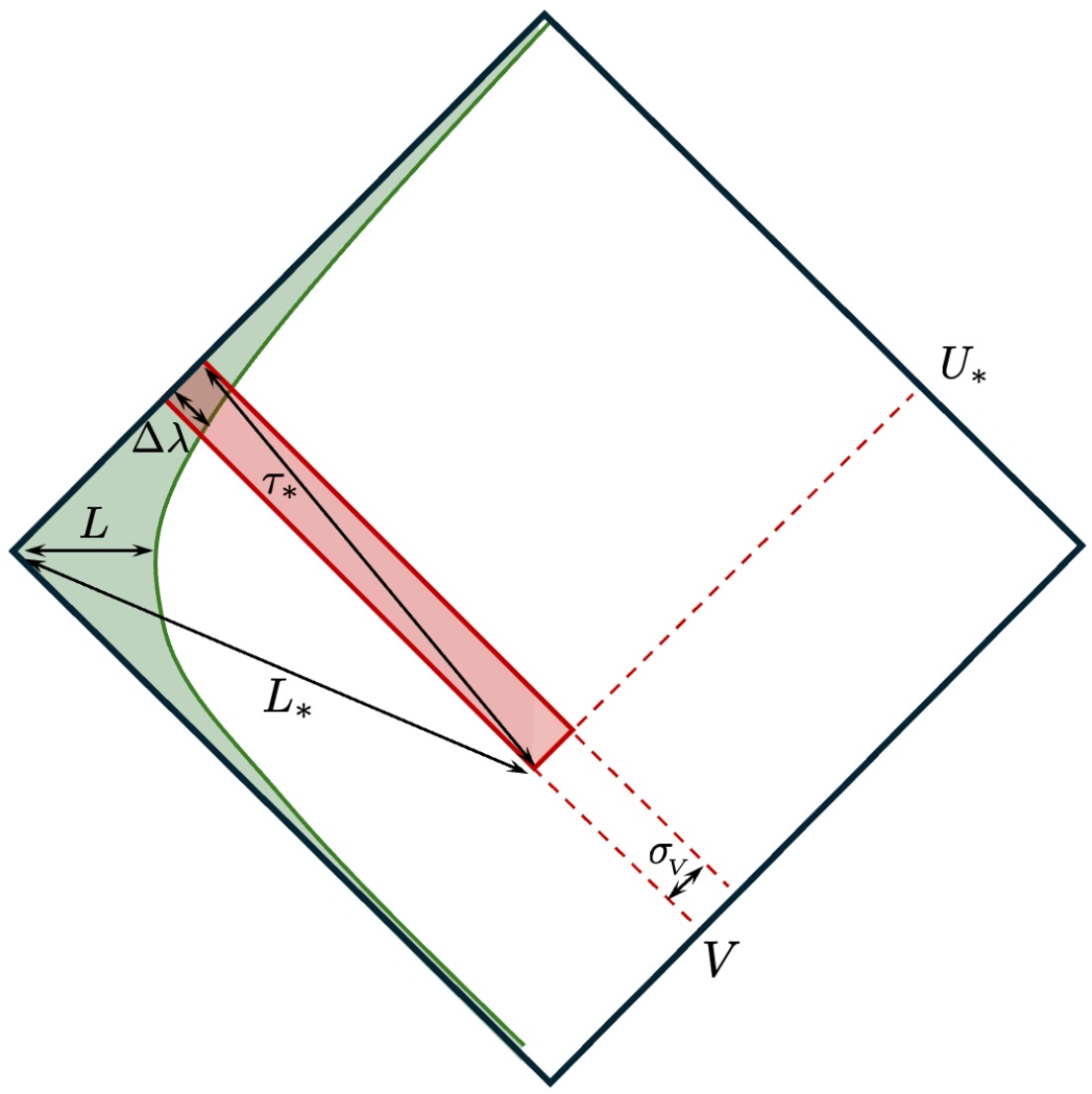}
    \caption{The setup and definition of the various physical quantities.}
    \label{Fig:set-up}
\end{figure}

To convert the variance in $\Delta \lambda$ to an estimate of the quantum width $L$ of the horizon, we make use of the local Rindler geometry near the black hole. This leads to the relation 
\begin{equation}
    L^2 = V \Delta \lambda
    \label{Deltalambda-to-L}
\end{equation}
where $V$ is the Kruskal coordinate near the horizon; see figure \ref{Fig:set-up}.
In general, the result can also depend on the affine time $U_*$ at which the experiment begins. This is related to the proper distance $L_*$ from the horizon at which the experiment begins via
\begin{equation}
     L_*^2 \equiv VU_*
\end{equation}
as can be seen from figure \ref{Fig:set-up}. We will later choose $L_*$ to be the location where the near-horizon Rindler approximation to the metric breaks down, but the observable is well-defined for any choice of $L_*$.

In perturbative gravity, fluctuations in the spacetime geometry are small at leading order, because they are proportional to Newton’s constant $G_N$. The physical set-up involves several other length scales, such as the black hole radius, affine distance to the horizon and the spatial and temporal resolution of the measurement. The measured quantum width of the horizon can in principle depend on all these quantities and thus may involve dimensionless ratios of these relevant length scales.  The goal of this paper is to determine the dependence of the quantum width $L$ on all of these length scales, while neglecting order one constants.

In Figure \ref{Fig:set-up} we have depicted the setup of our problem. In our calculations, we find that the size of the quantum width depends on the time scale $\tau_*$ associated with the experiment, as well as the transversal resolution $\sigma_\perp$ of the experiment.  We assume that the lightcone time $V$ at which the signal is sent towards the black hole is determined with resolution $\sigma_V$. The time scale $\tau_*$ is then defined via the relation
\begin{equation}
    \tau_*^2 = \sigma_V U^*
\end{equation}
and represents the proper time between the moment the light probe sends its signal and the moment a nearby lightray at $V+\sigma_V$ reaches the horizon, as indicated in Figure \ref{Fig:set-up}. The transversal resolution $\sigma_\perp$ determines the uncertainty in the value of the transversal momentum $k_\perp$ of our observable $\Delta\lambda$ along the horizon. To perform our calculations,  we  approximate the local near-horizon geometry by Rindler spacetime. The temporal and spatial resolutions are represented by smearing the observables over a small time interval or over a spatial region of size $\sigma_\perp$ along the horizon. 

We will begin by calculating the correlation functions of the variation $\Delta\lambda$ in the affine parameter as a function of the transverse momentum $k_\perp$. To regulate short-distance divergences, we smear our observables over an interval in the light-cone time $V$ with size $\sigma_V$. 
This leads to the following result valid for any spacetime dimension $D$
%\begin{equation}
%\boxed{    \bigl\langle \Delta \lambda_{\sigma_V} (k_\perp ) \Delta \lambda_{\sigma_V}(k_\perp') \bigr\rangle =   \delta^{D-2}(k_\perp + k_\perp') \times 
%    \begin{cases}
%   G_N/(\sigma_V^2 k_\perp^4) &\qquad  U_*\sigma_V k^2_\perp \gg 1 \\[1mm]
%   G_N U_*^2\left| \log (U_*\sigma_V k^2_\perp) \right| & \qquad  U_*\sigma_V k^2_\perp \ll 1
%    \end{cases}}
%\end{equation}
\begin{equation}
\boxed{   {1\over U_*^2} \bigl\langle \Delta \lambda_{\deltatau} (k_\perp ) \Delta \lambda_{\deltatau}(k_\perp') \bigr\rangle =   \delta^{D-2}(k_\perp + k_\perp') \times 
 \begin{cases}
 G_N/(\deltatau^4 k_\perp^4) &\qquad  \deltatau k_\perp \gg 1 \\[1mm]
G_N \left| \log (\deltatau k_\perp) \right| & \qquad  \deltatau k_\perp \ll 1
\end{cases}}
\end{equation}
Here we distinguish two regimes depending on the relative size of the transverse momentum compared to the time scale $\tau_*$.

These correlation functions can be converted to an expression for the quantum width $L$ of the horizon by integrating the transversal momentum with a gaussian measure with spatial resolution $\sigma_\perp$ and using the formula (\ref{Deltalambda-to-L}). When $\sigma_\perp$ is much larger than $\tau_*$ we find a universal answer 
\begin{equation}
   % {1\over U_*^2} \bigl \langle \, \Delta \lambda^2_{ \sigma_\perp, \deltatau} \bigr \rangle 
\boxed{   {1\over L_*^4} \bigl \langle \,(L^2_{\sigma_\perp, \deltatau})^2 \,\bigr \rangle
    =   
       {G_N \over \sigma_\perp^{D-2}} \: \log\left( {\sigma_\perp \over \deltatau} \right) \qquad\qquad \mbox{for} \quad \sigma_\perp \gg \deltatau}
\end{equation}
In the opposite regime, when the spatial resolution is much smaller than the temporal resolution, the quantum width also depends on the spacetime dimension $D$. In dimensions lower than six, the quantum width turns out to be entirely determined by the time resolution $\tau_*$ of the measurement, while in higher dimensions it also involves the transverse resolution $\sigma_\perp$. Our results split into three cases depending on the dimension $D$ 

%\section{Beyond the near-horizon approximation}
%There is an important loose end in the previous analysis. We calculated everything in the near-horizon limit, but IR divergences appeared, which we cut off at a proper distance of order $r_s$ away from the black hole horizon. As Temple He pointed out, it is unclear what the precise role of the black hole is in our computation. 
%In this section, we want to calculate the quantum width in the full black hole background. We hope to see that here our calculations are not IR divergent, and that they approximately reproduce the computations from the previous section, thereby justifying the cutoff prescription.
%For simplicity we calculate only the null definition of the horizon width. 

%\bc{Can be moved elsewhere}

%$$
%\left({\Delta \lambda\over U^*}\right)^2 = {G_N \over \sigma_\perp^{D-2}}\, %f\!\left({U^* \sigma_v\over \sigma_\perp^2}\right)
%$$

%\paragraph{Rindler Space:}
%If we calculate the fluctuations of a smeared observable in transverse momentum space, we find\footnote{Throughout the paper, we neglect order one constants.}

\begin{equation}
   % {1\over U_*^2} \bigl \langle \, \Delta \lambda^2_{ \sigma_\perp, \deltatau} \bigr \rangle 
\boxed{     {1\over L_*^4} \bigl \langle \,(L^2_{\sigma_\perp, \deltatau})^2 \,\bigr \rangle
    =   \begin{cases}
{\displaystyle{G_N  \over \deltatau^{D-2}} }&  
%\qquad %\sigma_\perp \ll  \deltatau \quad \ \mbox{and} 
%\quad \ 
(D< 6)
     \\[3mm]
      {\displaystyle{G_N  \over \deltatau^4}\: \log \left({\sigma_\perp\over \deltatau }\right)}&  %\qquad %\sigma_\perp \ll  \deltatau \quad \ \mbox{and} 
      %\quad \ 
      (D= 6) \\[3mm]
       {\displaystyle{ G_N \over \deltatau^4 \sigma_\perp^{D-6}}}& %\qquad 
       %\sigma_\perp \ll \deltatau \quad\  {\mbox{and}}
       %\quad \ 
       (D> 6)
    \end{cases}\qquad\qquad\quad \mbox{for} \quad \sigma_\perp \ll \deltatau}
\end{equation}
So far the equations are valid for black hole as well as Rindler horizons. As such, they do not depend on any specific properties of the black hole. It is perhaps surprising that the quantum width $L$ seems to depend on the IR scale $L_*$. In the black hole context, the value of the scale $L_*$ is naturally identified with the scale at which the black hole geometry starts to deviate from Rindler spacetime. In dimensions higher than six the IR dependence drops out if one replaces the time scale $\tau_*$ by the resolution $\sigma_t$ in the Schwarzschild time.
Their relation is given by
\begin{equation}
T\sigma_t = {\sigma_V\over V}%= \tau_*^2\over L_*^2 
=\left({\tau_*\over L_*}\right)^2
\end{equation}
where $T$ is again the Hawking temperature. 
The dependence on the IR scale and spacetime dimension $D$ are both removed by computing the fluctuations in the spherical harmonics of the quantum width as a function of the resolution $\sigma_t$ in the Schwarzschild time $t$. This leads to the following result, which entirely depends on natural length scales associated with the black hole and is valid in any dimension
\begin{equation}
\boxed{ {1\over r_s^4}\bigl \langle (L^2_{\ell, \sigma_t})^2 \bigr\rangle =   {G_N \over r_s^{D-2}}{1 \over \ell^4 T^2\sigma_t^2 }} \qquad \qquad \mbox{for}\qquad  {1 \over T\sigma_t \ell^2} \ll  {L_*^2 \over r_s^2}
\end{equation}
It is a striking fact that this universal result, which is obtained from a leading order perturbative analysis, looks practically the same as the naive estimate (\ref{thermo-estimate}) based on black hole thermodynamics. In fact, if we choose the time resolution $\sigma_t$ to be equal to $r_s/\ell^2$, the results are identical. 

This paper focuses on spacetime dimensions $D>3$. The case of $D=3$, where the quantum width can be calculated more precisely for BTZ black holes, is addressed in the parallel publication \cite{btz}. We thank Upamanyu Moitra for extensive discussions and collaboration.
%This value for $\sigma_t$ generally falls outside the validity range, so even for small values of $\ell$ our result will be smaller than the thermodynamic estimate. However, the total quantum width $L$ is obtained by summing over all values of $\ell$, and hence may exceed this estimate. 

%\bc{outline of paper}

\paragraph{Outline of the paper.}
We will start in section 2 by computing the correlation functions of the fluctuations $\Delta\lambda$ in the affine parameter distance to the Rindler horizon in flat Minkowski space. This calculation serves as a preparation for the analogous calculation for black hole horizons.  In section 3 we derive an expression for the fluctuation $\Delta\lambda$ of the affine distance to the black hole horizon in terms of the metric perturbations around a spherically symmetric black hole geometry. The calculations of the correlation functions of the fluctuations $\Delta \lambda$ are performed in section 4 and lead to the given expressions in transversal momentum space and angular momentum modes. In section 5 we translate these results to an estimate of the quantum width $L$ of the horizon. Finally, we discuss the thermodynamic interpretation in section 6. 

\paragraph{Related work.} In addition to the work mentioned above, we want to mention a number of other closely related papers. Previous work that builds on Marolf's quantum width \cite{Marolf2003} includes \cite{Hu:2006dd, Hu:2006jc}. These works estimate the backreaction of matter on the black hole horizon, a one-loop effect, while here we calculate the leading tree-level effect.

In addition, there have been a number of interesting calculations of fluctuations of causal diamonds more generally, including \cite{Verlinde:2019ade, Verlinde:2019xfb, Zurek:2020ukz, Banks:2021jwj, Zhang:2023mkf, Gukov:2022oed, Banks:2023wua, Ciambelli:2025fbo, Ciambelli:2026pwi, Fransen:2025npa, He:2025hag, Banks:2025erx, Freidel:2026hed}. The analogous quantity for the cosmological horizon was analyzed in \cite{Aalsma:2025bcg}. A perturbative calculation of interferometer-type observables in flat space \cite{Carney:2024wnp} found fluctuations at the Planck scale. Quantum gravity  fluctuations of timelike and null geodesics were calculated in \cite{Bak:2022oyn, Bak:2023wwo}. Enhanced fluctuations of flat space observables that are similar in spirit to our observables were found within perturbative quantum gravity by Bonga and Khavkine \cite{Bonga:2013uha}.

\section{Fluctuations of the Rindler Horizon}
\label{sec:rindler}
It is worthwhile to begin with a careful treatment of the fluctuations of the Rindler horizon in order to understand subtleties about infrared divergences, etc., in a situation where the calculations are technically simple. 

We work in arbitrary dimensions $D>3$, because some infrared issues are dimension dependent. We take the metric to be
\begin{equation}
ds^2 = - 2 dU dV + dy_\perp^2 + h_{ab} dx^a dx^b
\end{equation}
with the metric perturbation general at this point.

We assume the metric perturbation goes to 0 at infinity. The codimension-2 Rindler horizon is defined in the perturbed metric as the intersection of the future lightcone of $(U=-\infty,V=0, \vec y_\perp)$ with the backward lightcone of $(U=0, V = \infty, \vec y_\perp)$. The full Rindler horizon is the union of the two lightsheets. We will be interested in the fluctuations of the future Rindler horizon.

We want to calculate the fluctuations in the affine parameter of a `radial' null geodesic from $\mathscr{I}_-$ to the Rindler horizon. The overall normalization of the affine parameter is arbitrary; we choose it to be normalized such that
\begin{equation}
d\lambda = d U \end{equation} asymptotically.

We will first calculate the fluctuation in affine parameter in any arbitrary perturbed metric, then calculate the fluctuations in this quantity. This calculation can of course be done in any gauge, but it is very convenient to use a partial gauge fixing in which the perturbation satisfies
\begin{equation}
h_{UU} = h_{VV}=0
\end{equation} 
so that 
\begin{equation} \label{eqn:metconf}
    ds^2 = - 2 e^\phi dU dV + 2 h_{Ui} dU dy^i + 2 h_{Vi} dV dy^i + h_{ij} dy^i dy^j
\end{equation}
The location of the future Rindler horizon is given by the backward lightcone of  $(U=0, V = \infty, \vec y_\perp)$. To linear order in the perturbations, in this partially fixed gauge,  the lightcone is simply
\begin{equation}
    U = 0
\end{equation}
 The null geodesics are given by solutions to the equation of motion of the action
\begin{equation}
    S = \int d\lambda g_{a b} \dot x^a \dot x^b
\end{equation}
where $\dot x$ denotes derivative with respect to the affine parameter $\lambda$. Similar to the equation for the Rindler horizon, one can show that, while the mixed components $h_{Vi}$ and $h_{Ui}$ are important for the motion in the transverse directions $y^i$, they do not affect the affine parameter to the horizon to linear order, so we can simply use the action
\begin{equation}
S = \int d\lambda e^\phi \dot U \dot V
\end{equation}
The equation of motion gives
\begin{equation}
    {d \over d \lambda}(e^\phi \dot U) = \dot U \dot V \partial_V e^\phi
\end{equation}
Since $\dot V= 0$ along the geodesic of interest to linear order, the equation is just
\begin{equation} 
e^\phi \dot U = 1
\end{equation}
where we have chose the normalization $\dot U = 1$ at infinity, where the fluctuations $\phi$ are assumed to vanish. Therefore, the affine parameter to the horizon is 
\begin{equation} 
\lambda = \int_{-\infty}^0 dU e^{\phi}
\end{equation}
Expanding for small perturbations, the fluctuations in affine parameter are given by
\begin{equation} \label{eqn:dellambdadefn}
    \Delta \lambda(V, y^i) = \int_{-\infty}^0 dU \phi(U, V, y^i)
\end{equation}

As we will find below, this observable
$\Delta \lambda$ can exhibit IR divergences
when integrating all the way to $U = -\infty$.
We will therefore also need to incorporate an IR 
cutoff in the $U$ integral to parameterize when
this IR sensitivity becomes important.  This 
raises the question of whether the choice 
of such an IR cutoff is gauge invariant.  

The conformal gauge choice employed when writing
the metric ansatz (\ref{eqn:metconf}) can be viewed
as determining the $U$ coordinate by following
lightrays fired from $\mathscr{I}_+$, while
the $V$ parameter comes from lightrays fired from
$\mathscr{I}_-$.  This fixes these coordinates sufficiently
well to linear order that we can view these 
coordinates as well as fields depending on them
as gauge invariant.  Hence any IR cutoff function
evaluated in the conformal coordinate system
here will lead to a gauge-invariant observerable 
at this order.  

A related point involves the choice of how to 
normalize the affine parameter.  The observable
defined by (\ref{eqn:dellambdadefn}) involves a 
choice to fix the affine parameter by fixing 
the inner product of the ingoing lightray with 
a null ray tangent to $\mathscr{I}_-$.  We could instead
choose to fix the inner product with a null lightray
coming from $\mathscr{I}_+$ at coordinate $U_*$
that intersects
our geodesic.  This choice would modify the observable
to be
\beq\label{eqn:scri+norm}
\Delta\lambda = \int_{-\infty}^ 0 dU[\phi(U,V) - \phi(U_*, V)].
\eeq
This extra term $\phi(U_*,V)$ leads to different results
for the two-point function in the observable, which 
is not surprising since this different choice 
of normalization simply defines a different 
gauge-invariant observable.  We include 
some expressions involving this modified observable 
in the appendix \ref{app:scri+}, but for the remainder 
of this section we focus on the $\mathscr{I}_-$
normalization corresponding to (\ref{eqn:dellambdadefn}).

% \AJS{incorporating this discussion here}
% In the previous sections we have been interested in the fluctuations in the affine parameter from infinity to the Rindler horizon. Suppose instead we want to do a more localized experiment. Then we need to be careful to choose the starting point in  a gauge invariant way. The most natural starting point for our experiment is to intersect null rays from $\mathscr{I}_-$ and $\mathscr{I}_+$. In our gauge, to linear order, this simply means picking an initial $(U_*, V)$ where our null ray begins.

% We also need to choose how to normalize the affine parameter. We can either use the inner product of the ingoing null ray with the ray heading towards $\mathscr{I}_+$, or we can normalize it by fixing the normalization at $\mathscr{I}_-$. These are two different choices. Normalizing with respect to $\mathscr{I}_+$ yields
% \begin{equation}
%     \Delta \lambda = \int_{U_*}^0 dU [\phi(U, V) - \phi(U_*, V)]
% \end{equation}
% while normalizing with respect to $\mathscr{I}_-$ gives 
% \begin{equation}
%     \Delta \lambda = \int_{U_*}^0 dU \phi(U, V) 
% \end{equation}

\subsection{Calculating the fluctuations}

In order to calculate the fluctuations we need to know the correlators in this partially specified gauge. By for example transforming the de Donder gauge propagator to this `conformal' gauge, we find, up to order one constants which we neglect throughout 
\begin{equation} \langle \phi(x) \phi(y) \rangle = G_N G_{\rm scalar} (x, y)
\end{equation}
The correlator is then given by
\begin{equation}
\langle \Delta \lambda(V, y^i) \Delta \lambda( V', y^{i'}) \rangle = G_N \int_{-\infty}^0 dU  \int_{-\infty}^0 dU' G_\text{scalar}(U, V, y^i; U', V', y^{i'})
\end{equation}
This can be calculated in position space,
\begin{equation}
\langle \Delta \lambda(V, y^i) \Delta \lambda( V', y^{i'}) \rangle = G_N \int_{-\infty}^0 dU  \int_{-\infty}^0 dU' {1 \over [(\Delta y)^2 - 2 (U - U') \Delta V - i \epsilon \delta t]^{(d-2)/2}}
\end{equation}
We are particularly interested in possible IR divergences, since this expression clearly has the possibility to behave badly at $U \to -\infty$ in some dimensions.

We can parameterize the IR-sensitive behavior by explicitly 
introducing an IR-cutoff into the $U$-integral defining
$\Delta\lambda$, and then analyzing the sensitivity 
of the answer to this cutoff.  Hence, the IR-regulated
observable will be given by
\beq \label{eqn:lambdaIR}
\Delta\lambda_\Lambda = \int_{-\infty}^0 dU \Lambda(U) 
\phi(U,V, y^i),
\eeq
where $\Lambda(U)$ is a cutoff function satisfying $\Lambda(0) = 1$
and that goes to zero sufficiently rapidly as 
$U\rightarrow-\infty$.
One natural choice in position
space is a hard cutoff $\Lambda(U) = \Theta(\irs + U)$.  Another
choice that is more convenient in momentum space 
is an exponential cutoff $\Lambda(U) = e^{\frac{U}{\irs}}$.  
We will use the exponential cutoff in the analysis below.

With this IR regulator, the observable has a convenient 
momentum space expression
\begin{equation}
    \Delta \lambda(k_V, k_\perp) = \int_{-\infty}^0 dU \int dk_U e^{i (k_U-\frac{i}{\irs}) U} \phi(k_U, k_V, k_\perp).
\end{equation}
Performing the $U$ integral, we obtain
\begin{equation}
    \Delta \lambda(k_V, k_\perp) = \int dk_U {1 \over i (k_U - \frac{i}{\irs})} \phi(k_U, k_V, k_\perp).
\end{equation}
Using the momentum-space Wightman function
\begin{equation}
    G_{\rm scalar}(k,k') = \delta^D(k+ k') \Theta(k_U) \delta(k^2).
\end{equation}
we get the correlator
\begin{equation}
    \langle \Delta \lambda (k_V, k_\perp) \Delta \lambda(k_V', k_\perp') \rangle = G_N \delta^p(k_\perp + k') \delta(k_V + k_V') \int_0^\infty  dk_U {1 \over k_U^2 + \frac{1}{\irs^2}} \delta (2 k_U k_V - k_\perp^2)
\end{equation}
Using the delta function to do the integral, we have finally
\begin{equation} \label{eqn:dellambdak}
     \langle \Delta \lambda (k_V, k_\perp) \Delta \lambda(k_V', k_\perp') \rangle = G_N \delta^p(k_\perp + k') \Theta(k_V) \delta(k_V + k_V'){2 k_V \over k_\perp^4 + \frac{4k_V^2}{\irs^2}}
\end{equation}
Here and in the following we denote the horizon dimension by $p\equiv D-2$. 
The possible IR issues at $U = \infty $ that we saw in position space would now be expected to appear at $k_U = 0$. The on-shell delta function sets
\begin{equation}
    k_U =  k_\perp^2/(2 k_V)
\end{equation}
so $k_U \to 0$ is only possible if $k_V \to \infty$ or $k_\perp \to 0$. We will prevent $k_V \to \infty$ by smearing in $V$ by an amount $\sigma_V$,  and discuss $k_\perp \to 0$ momentarily.  Going back to position space in the $V$ coordinate, we have
\begin{align}
    \langle \Delta \lambda (V, k_\perp)  \Delta \lambda (V', k_\perp') \rangle &=  G_N \delta^p(k_\perp + k'_\perp ) \int_0^\infty dk_V {2 k_V \over {k_\perp^4+ \frac{4 k_V^2}{\irs^2}}} e^{i k_V (\Delta V+ i \epsilon)} 
    \nonumber \\
    &= 
    G_N \delta^p(k_\perp + k_\perp')\frac{\ell_{\irs}^2}{4} F(a), \label{eqn:Dlk}
\end{align}
where
\beq
a = \frac12\irs
    k_\perp^2(\Delta V + i \epsilon)
\eeq
and the function $F(a)$ is given by
\beq
F(a) = e^{-a}(i\pi -2 \Shi(a)) + 2 \cosh(a) E_1(a),
\eeq
with $\Shi(a)$ the hyperbolic sine integral function
and $E_1(a)$ the exponential integral function.  

At large arguments, $F(a)$ has an asymptotic 
expansion going like $F(a) \sim \frac{-2}{a^2}$.  
Applying this to (\ref{eqn:Dlk}) for $\irs k_\perp^2
(\Delta V + i \epsilon) \gg 1$, the correlation
function in this regime becomes independent 
of the IR regulator, approaching 
\beq \label{eqn:k4V}
 \langle \Delta \lambda (V, k_\perp)  \Delta \lambda (V', k_\perp') \rangle
 \rightarrow
 G_N \delta^p(k_\perp+k_\perp') \frac{-2}{k_\perp^4 (\Delta V + i \epsilon)^2}
\eeq

The factor of $i \epsilon$ in (\ref{eqn:Dlk})  ensures that the 
integral converges even in the $\Delta V \rightarrow 0$
limit, so that the final answer is well defined 
as a distribution in the $\Delta V$ coordinate.  
Keeping $\epsilon$ small but finite is roughly 
equivalent to smearing the observables in $V$,
which produces operators with finite fluctuations.  
If we consider the fluctuations of a smeared operator
with smearing width in $V$ of order $\sigma_V$, we have 
\begin{equation} \label{eqn:sigVsmear}
    \langle \Delta \lambda_{\sigma_V} (V,  k_\perp)  \Delta \lambda_{\sigma_V} (V,  k_\perp') \rangle \sim   \delta^{p}(k_\perp + k_\perp'){G_N \over (\sigma_V)^2 k_\perp^4}
\end{equation}
Here it is clear that, aside from possible issues at $k_\perp \to 0$, there are no IR issues. This expression
remains valid for $k_\perp^2 \irs \sigma_v \gg 1$,
while for $k_\perp^2 \lesssim \frac{1}{\irs \sigma_v}$
it is modified to suppress the contributions from
very small $k_\perp$.  Hence when using this expression,
there is an effective cutoff on $k_\perp$ around
the scale $\frac{1}{\sqrt{\irs \sigma_v}}$.  

The regime $k_\perp \rightarrow 0$ instead involves
the small argument expansion of $F(a)$, given 
by\footnote{Note that this formula assumes that $\Im(a) >0$ (or $\Im(a)\rightarrow 0^+$) as 
given by the $i\epsilon$ prescription above, and from
this one can verify that for real $x$, 
$\Im(F(x+i \epsilon)) = 
-\Im(F(-x+i\epsilon))$.} 
\beq
F(a) \overset{a\ll 1}{=} -2\log a -2\gamma_E + 
i(\pi-a) +\ldots .
\eeq
Hence for $\irs k_\perp^2 \Delta V \ll 1$, the correlator
behaves as 
\beq
\langle \Delta \lambda (V, k_\perp)  \Delta \lambda (V', k_\perp') \rangle \rightarrow 
-G_N \delta^p(k_\perp + k_\perp') 
\frac{\irs^2}{2} \log\left(\frac{\irs k_\perp^2
(\Delta V + i \epsilon)}{2}\right) 
\eeq
with quadratic and logarithmic sensitivity to the 
IR scale $\irs$.  

It is worthwhile to understand this directly in position space. Using the position space propagator and 
the exponential regulator, we have 
\begin{equation}
\langle \Delta \lambda(V, y^i) \Delta \lambda( V', y^{i'}) \rangle = \frac12 G_N \int_{-\infty}^0 d\bar U \int_{-\bar U}^{\bar U} d(\Delta U) {e^{\frac{\bar{U}}{\irs}} \over [(\Delta y)^2 - 2 \Delta U \Delta V - i \epsilon \delta t]^{(d-2)/2}}
\end{equation}
Performing the $\Delta U$ integral gives
\begin{align}
&\langle \Delta \lambda(V, y^i)  \Delta \lambda( V', y^{i'}) \rangle = 
 {G_N \over \Delta V + i \epsilon}\frac{1}{2(d-4)} \times
 \nonumber \\
 &\quad\int_{-\infty}^0 d\bar U  e^{\frac{\bar U}{\irs}} \left[{1 \over [(\Delta y)^2 - 2 \bar U \Delta V - i \epsilon \delta t]^{(d-4)/2}} -  {1 \over [(\Delta y)^2 + 2 \bar U \Delta V - i \epsilon \delta t]^{(d-4)/2}}\right]
\end{align}
In sufficiently low dimensions, the terms in brackets
lead to a divergent integral in the absence of the IR 
cutoff provided by $e^{\frac{\bar U}{\irs}}$.  However,
it clearly is convergent for $d>6$, and in this case the dependence on the IR scale is negligible in the large
$\irs$ limit.  This is precisely the regime
where the inverse Fourier transform of 
(\ref{eqn:k4V}) is well-defined. 

Finally, it is worth displaying the expressions
for the fluctuations of the affine parameter after
smearing the observable in the transverse directions,
rather than analyzing the exact Fourier modes $k_\perp$.  
We can use a Gaussian smearing for simplicity,
and employ the approximate expression 
(\ref{eqn:sigVsmear}) to get
\beq
\langle (\Delta\lambda_{\sigma_V\sigma_\perp})^2\rangle
\sim
\frac{-2 G_N}{\sigma_V^2} \int d^pk_\perp e^{-\sigma_\perp^2k_\perp^2}
\frac{1}{k_\perp^4}.
\eeq
Remembering that the 
IR region is cut off at
$|k_\perp| \sim \frac{1}{\sqrt{\irs\sigma_V}}$,
we find that it
is sensitive to the IR cutoff when $p\leq 4$,
corresponding to spacetime dimension $D\leq 6$.  
The result scales as
\begin{equation}
       \langle (\Delta \lambda_{\sigma_V \, \sigma_\perp})^2 \rangle \sim \begin{cases} {\displaystyle G_N \over \sigma_V^2 \sigma_\perp^{p-4}} & D>6\\
       {G_N \over \sigma_V^2} \log(\sigma_\perp (\sigma_V \irs)^{-\frac12}) & D = 6 \\
       {G_N \irs^2 \over {(\sigma_V\irs)^{\frac{p}{2}}}} & D < 6
       \end{cases}
\end{equation}

% \paragraph{Position space expression}
% One may be interested in calculating the fluctuations of the affine parameter averaged with some smearing function over the transverse dimensions, rather than in Fourier space. Taking the smearing to be Gaussian for simplicity, we have
% \begin{equation}
%     \langle (\Delta \lambda_{\sigma_V \, \sigma_\perp})^2 \rangle = {G_N \over \sigma_V^2} \int d^p k_\perp e^{- \sigma_\perp^2 k_\perp^2}{1 \over k_\perp^4}
% \end{equation}
% This integral is IR divergent in $p\leq 4$, corresponding to spacetime dimensions $d \leq 6$. The result of the integral is
% \begin{equation}
%        \langle (\Delta \lambda_{\sigma_V \, \sigma_\perp})^2 \rangle = \begin{cases} {\displaystyle G_N \over \sigma_V^2 \sigma_\perp^{p-4}} & d>6\\
%        {G_N \over \sigma_V^2} \log(\Lambda_{\rm IR} \sigma_\perp) & d = 6 \\
%        {G_N \Lambda_{\rm IR}^{4-p} \over \sigma_V^2} & d < 6
%        \end{cases}
% \end{equation}
% So if we want to do experiments in $d< 6$, we need to determine what physical quantity determines the IR  cutoff. In addition, we want to use these results to analyze the near horizon region of black holes, where it is also important to determine the sensitivity to IR physics. To understand this better we now consider an IR finite experiment. 

It is also worth considering how sensitively
this answer depends on the details of the IR cutoff.  
The results above were obtained using an
exponentially decaying IR cutoff to define
the IR regulated observable in (\ref{eqn:lambdaIR}).  
Another natural choice is to use a hard cutoff
in the $U$ parameter, corresponding to 
$\Lambda(U) = \Theta(U+U_*)$.  Following the same
steps as before, in this case the momentum-space
correlator becomes
\beq \label{eqn:dellambdakU}
\langle\Delta\lambda(k_V, k_\perp), \Delta\lambda(k_V',
k_\perp')\rangle = 
G_N \delta^p(k_\perp+k_\perp') \delta(k_V + k_V')
\Theta(k_V) \frac{8 k_V\sin^2\left(\frac{k_\perp^2 U_*}{4 k_V}\right) }{k_\perp^4}
\eeq
In the IR regime corresponding to $\frac{k_\perp^2 U_*}{k_V} \ll 1$, the quantity multiplying the 
$\delta$ and $\Theta$ functions approaches
\beq
G_N\frac{U_*^2}{2 k_V},
\eeq
while the same limit of the exponentially regulated
expression (\ref{eqn:dellambdak}) gives
\beq
G_N \frac{\irs^2}{2 k_V}.
\eeq
Hence we see the two correlation functions agree
in this limit upon identifying $U_*$ with $\irs$.  

In the opposite regime $\frac{k_\perp^2 U_*}{k_V} \gg 1$, 
the correlator (\ref{eqn:dellambdakU}) is a rapidly
oscillating function of $k_V$.  However, smearing the
answer over a small window in $k_V$ causes the oscillations
to average to their mean, and allows $\sin^2$ to 
be replaced by $\frac12$.  For a smearing width
in $V$ characterized by $\sigma_V$, this yields 
the correlator
\beq \label{eqn:lambdaUstar}
\langle\Delta \lambda_{\sigma_V}(k_\perp)
\Delta \lambda_{\sigma_V}(k_\perp')\rangle
\sim \delta^p(k_\perp + k_\perp ') \frac{2 G_N}{k_\perp^4
\sigma_V^2},
\eeq
which again agrees with (\ref{eqn:sigVsmear}).\footnote{If
tracking factors of $2$, 
the expression (\ref{eqn:lambdaUstar}) is 
actually twice as big as (\ref{eqn:sigVsmear}).  
This appears to come from an additional UV contribution
to the correlator due to the hard cutoff at 
$U_*$, which is therefore just an effect
associated with the uncertainty of the starting
point of the experiment.  For this reason,
the exponential cutoff appears more reliable
for considerations involving fluctuations of the 
horizon. }
The remaining expressions are similar to those
obtained with the exponential cutoff, so we conclude
by summarizing the behavior in different 
regimes, after smearing 
in both $V$ and the transverse direction $y^i$: 

\begin{equation}
    \langle (\Delta \lambda_{\sigma_V, \sigma_\perp} (U_*) )^2 \rangle \sim   \begin{cases}
       {\displaystyle{ G_N \over \sigma_V^2 \sigma_\perp^{D-6}}}& \sigma_\perp^2 \ll U_* \sigma_V \ \ {\rm and} \ \ D>6 \\
       {\displaystyle \frac{G_N}{\sigma_V^2}
       \log\left(\frac{\sigma_\perp^2}{\sigma_V U_*}\right)} & \sigma_\perp^2 \ll U_*\sigma_V\ \ 
       {\rm and} \ \ D = 6 \\
     {\displaystyle{G_N U_*^2 \over (U_* \sigma_V)^{(D-2)/2}} }&  \sigma_\perp^2 \ll U_* \sigma_V \ \ {\rm and} \ \ D<6 \\
       {\displaystyle{G_N U_*^2 \over \sigma_\perp^{D-2}}}  \log {\sigma_\perp^2 \over U_* \sigma_V} & \sigma_\perp^2 \gg U_* \sigma_V 
    \end{cases}
\end{equation}

\bigskip

\section{Fluctuations of the Black Hole Horizon}
We begin from the Eddington-Finkelstein coordinates, and do not yet make any gauge choices. The metric is 
\begin{equation}
ds^2 = 2(1 + h_{vr}) dv dr - (f(r) - h_{vv}) dV^2 + h_{rr} dr^2 + \dots
\end{equation}
We have not written the angular parts of the metric or the mixed perturbation components, such as $h_{v\theta}$, because these do not contribute at linear order to the radial null geodesics we consider.

\subsection{Perturbed Event Horizon.} We first calculate the location of the perturbed future event horizon. In the background, the horizon is given by $r= r_s$, with $r_s$ defined by $f(r_s) = 0$. In the perturbed geometry, the event horizon is still null. We expand near the horizon, expanding $f(r) \approx f_0' (r - r_s)$. We also use that since $r$ is constant on the unperturbed horizon, $dr$ is small. Expanding to linear order in perturbations, we have
\begin{equation}
0 = 2 dv dr - [f'_0 (r - r_s) - h_{vv}]dv^2 
\end{equation}
The future horizon has $dV \neq 0$, so it obeys
\begin{equation}
dr = {1 \over 2}[f'_0 (r - r_s) - h_{vv}] dv
\end{equation}
This can be rewritten as
\begin{equation}
{d(r - r_s)  \over dv} - {f'_0 \over 2} (r - r_s) = - {1\over 2} h_{vv}
\end{equation}
We want to impose the boundary condition that the horizon is unperturbed at late time, $r \to r_s$ as $v \to \infty $. The solution obeying this boundary condition is
\begin{equation}
r(v) - r_s = {1 \over 2} \int_v^\infty  dv' h_{vv}(v', r_s) e^{-{f'_0 \over 2} (v' - v)}
\end{equation}
This is an integral over a part of the future horizon. The expression looks even simpler when written in terms of the Kruskal $v$. Using the formula from Appendix A,
\begin{equation}
V = r_s e^{v f'_0/2}
\end{equation}
and changing variables gives
\begin{equation}
r(V) - r_s = {f'_0 \over 4} V \int_V^\infty dV' h_{VV}(V', r_s)
\end{equation}
This can also be rewritten in terms of the Kruskal $U$ coordinate\footnote{Note that at generic locations $h_{vv}^{EF}$ in the Eddington-Finkelstein coordinates is a combination of several components in the Kruskal coordinates. However, on the future horizon  $h_{vv}^{EF}$ is proportional to $h_{VV}^K$}. As reviewed in Appendix A, near the horizon 
\begin{equation}
- UV \approx {2 \over f_0'}(r-r_s)
\end{equation}
so in terms of the Kruskal $U$ coordinate the expression takes a simple form matching our Rindler calculations,
\begin{equation}
U(V) = - {1 \over 2} \int_V^\infty dV' h_{VV}(V', 0)
\end{equation}
The main motivation for using the Eddington-Finkelstein coordinates is that the ingoing null ray, which we calculate next, takes a simpler form in these coordinates.

\subsection{Perturbed Ingoing Null Geodesic}
The ingoing ray has constant $v$ in the background. In the presence of perturbations, keeping terms to linear order, it satisfies
\begin{equation}
0 = 2 dv dr + h_{rr} dr^2
\end{equation}
where we have used that $dv$ is of order $h$. Since $dr \neq 0$ the ingoing ray satisfies
\begin{equation}
dv = - {1 \over 2} h_{rr} dr
\end{equation}
For this null geodesic we also want to know the affine parameter and how it is perturbed. For this, we use the action that gives the equation of motion for geodesics in terms of affine parameter,
\begin{equation}
S = \int d\lambda g_{ab} \dot x^a \dot x^b
\end{equation}
Here $\dot x$ denotes derivative with respect to affine parameter $\lambda$. We also need to remember the constraint
\begin{equation}
g_{ab} \dot x^a \dot x^b = 0
\end{equation}
Once again, the motion in angular directions is small to this order and does not affect the motion or affine parameter in the $(r, t)$ plane. Keeping only terms up to quadratic order in the perturbations the action becomes
\begin{equation}
S = \int d\lambda \left[ 2(1 + h_{vr}) \dot v \dot r - f \dot v^2 + h_{rr} \dot r^2    \right]
\end{equation}
We are ultimately interested in solving for the affine parameter as a function of $r$ along the geodesic, so it is useful to rewrite the action thinking of $r$ as the independent variable and $\lambda$ as dependent variable. Using prime to denote differentiation with respect to $r$,
\begin{equation}
S = \int dr \left[ 2(1 + h_{vr}){v' \over \lambda'} - f{v'^2 \over \lambda'} + {h_{rr} \over \lambda'} \right]
\end{equation}
The $\lambda$ equation of motion is satisfied as long as the constraint is obeyed, so we just need the $v$ equation of motion. It reads
\begin{equation}
{d \over d r}[2(1 + h_{vr}) {1 \over \lambda'} - 2 f {v' \over \lambda'}] = 2 {v' \over \lambda'}\partial_v{h_{vr}} + {1 \over \lambda'} \partial_v h_{rr}
\end{equation}
In the background the affine parameter is simply proportional to the radial coordinate $r$,
\begin{equation}
\lambda' = {1 \over c}  + \dots
\end{equation}
with $c$ a constant that will be determined later by suitably normalizing the affine parameter at infinity. Since $\lambda'$ is constant at 0th order in the expansion, $\lambda''$ is first order. We can also use the previous result for the perturbed geodesic
\begin{equation}
v' = - {1 \over 2} h_{rr} 
\end{equation}
Combining these, the above equation can be written
\begin{equation}
-2 {\lambda'' \over \lambda'^2} + {2 \over \lambda'} \partial_r h_{vr} + {1 \over \lambda'}\partial_r(f h_{rr}) = {1 \over \lambda'} \partial_v h_{rr}
\end{equation}
Here we have used that $v$ is constant to 0th order, so the total $r$ derivative can be replaced by a partial $r$ derivative to 0th order.

Now using that $\lambda' = c$ to 0th order, we have
\begin{equation}
c \lambda'' =  \partial_r h_{vr} + {1 \over 2} \partial_r(f h_{rr}) - {1 \over 2} \partial_v h_{rr}
\end{equation}
Taking the boundary condition that the affine parameter and its derivative are unperturbed at the far initial point, labelled $R_0$, the perturbed affine parameter is
\begin{equation}
c  \lambda = r - {R_0} + \int_{R_0}^r dr' (r - r') \left[    \partial_r h_{vr} + {1 \over 2} \partial_r(f h_rr) - {1 \over 2} \partial_v h_{rr}  \right]
\end{equation}
where the first term on the right hand side is the background solution and the integral captures the perturbation.

\subsection{Affine parameter to the horizon}
 Now we are ready to combine the results of the previous subsections.
The affine parameter to the horizon is 
\begin{equation}
c \lambda_h = r_h - R_0 + \int_{R_0}^{r_s} dr (r_s - r) \left[    \partial_r h_{vr} + {1 \over 2} \partial_r(f h_{rr}) - {1 \over 2} \partial_v h_{rr}  \right]
\end{equation}
We are interested in the fluctuations in this quantity. Subtracting the background value and using the formula for the perturbed horizon location  gives
\begin{equation}
c \Delta \lambda_h = {1 \over 2} \int_v^\infty  dv' h_{vv}(v', r_s) e^{-{f'_0 \over 2} (v' - v)} + \int_{R_0}^{r_s} dr (r_s - r) \left[    \partial_r h_{vr} + {1 \over 2} \partial_r(f h_{rr}) - {1 \over 2} \partial_v h_{rr}  \right]
\end{equation}
Note that the first integral is along the horizon, while the second integral is along the ingoing null ray.

The $r$ derivative acts on the 2d trace $h$ of the metric perturbation,
\begin{equation}
    h \equiv 2 h_{vr} + f h_{rr}
\end{equation} 
We can simplify by integrating by parts in the second integral. We want to be careful about possible IR issues, so we keep track of all boundary terms. We have
\begin{eqnarray}
c \Delta \lambda_h = {1 \over 2} \int_v^\infty  dv' h_{vv}(v', r_s) e^{-{f'_0 \over 2} (v' - v)} + {1 \over 2}\int_{R_0}^{r_s} dr  \left[   h - (r_s - r) \partial_v h_{rr}  \right] 
 + {1 \over 2}\left[(r_s - r) h\right]_{R_0}^{r_s}
\end{eqnarray}
This is the final expression for the perturbation in the affine parameter to the horizon expressed in the Eddington-Finkelstein coordinates for the background. We have not chosen any gauge yet for the perturbations.

One straightforward but nontrivial check of this formula is that it is invariant under gauge transformations 
\begin{equation}
h_{ab} \to h_{ab} + D_{(a}\xi_{b)}
\end{equation}
as long as the diffemorphisms $\xi_a$ go to zero at the far endpoints.

\paragraph{Kruskal coordinates and Gauge Choice}
 
The expression can also be written in Kruskal coordinates, but in a general gauge this is somewhat messy. It is convenient to make the analogous gauge choice that we made in our analysis of the Rindler horizon,
\begin{equation}
h_{UU} = h_{VV}= 0 \ \ \ \ \ \ {\rm (gauge \ choice.)}
\end{equation}
Making use of the change of coordinates in the appendix, equations \eqref{k2ef} and \eqref{ef2k}, we have in this gauge
\begin{equation}
    h_{rr} = 0 \ \ \ \ \ \ \ \ h_{vv}(r_s)= 0 \ \ \ \ \ \ \ \ \ dr = {f \over f_0'U} dU \ \  {\rm for}\ V= {\rm const}
\end{equation}
With this gauge choice, the fluctuation in affine parameter becomes
\begin{equation}
    c \Delta \lambda = {1 \over 2} \left[(r_s - r) h \right]_{R_0}^{r_s} + {1 \over 2 }\int_{U_*}^0 dU {f \over f_0' U} h \ \ \ \ \ \ \ \  ({\rm Gauge}\ h_{UU}= h_{VV}=0)
\end{equation}
We further assume that the 2d trace $h$ is well-behaved at the horizon, and that it falls off at large $r$ fast enough that the boundary term can be neglected. With these assumptions and choices, we finally have the simple expression
\begin{equation}
    c \Delta \lambda = {1 \over 2 }\int_{U_*}^0 dU {f \over f_0' U} h \ \ \ \ \ \ \ \  ({\rm Gauge}\ h_{UU}= h_{VV}=0)
\end{equation}
where the limit of integration $U_*$ is given by change of coordinates from the $R_0$ specifying the starting point of the experiment; we will want to take $R_0 \to \infty$ later.

\section{Calculating the fluctuations}
Now that we have written the perturbation in the affine parameter in terms of the metric perturbation, would like to calculate its 2-point function perturbatively. An obstacle is that the graviton correlator in the Schwarzschild background cannot be written in closed form in any gauge, as far as we are aware. 

We will calculate the near-horizon part of the fluctuations, by taking the near-horizon limit of the formula above, and using known formulas for the graviton correlator in Rindler (= Minkowksi) spacetime. We will then check whether the near-horizon approximation is self-consistent. It is an interesting challenge for the future to perform the full black hole calculation, beyond this near-horizon limit.

We now use the near horizon approximation for $f$,
\begin{equation}
f \approx f_0' (r - r_s) = -{(f_0')^2 \over 2} UV \ \ \ \ ; \ \ \ \ 
r - r_s = -{UV f_0' \over 2} \ \ \ \ \ \  \ \ \; g_{UV} \approx -1
\end{equation}
to get
\begin{eqnarray}
c \Delta \lambda_h = -{f_0' V \over 4} \int_{U_*}^0 dU h = {f_0' V \over 2}\int_{U_*}^0 dU h_{UV}
\end{eqnarray}
We now choose the normalization $c$. We take the point of view that the normalization is fixed at the starting point of the experiment, far from the black hole where fluctuations are small. In the background, the affine parameter is proportional to the Schwarzschild radial coordinate $r$. We take the simplest choice and make the constant of proportionality 1,
\begin{equation}
    d\lambda = dr \ \ \ \ \rightarrow \ \ \ \ c=1\ .
\end{equation}
Note that in principle we could choose a $V$-dependent normalization of the affine parameter in order to cancel the $V$ dependence in the formula; however, this is not natural from the perspective of an experiment calibrated at large radius. Cancelling the $V$ dependence would require a time-dependent normalization at large $r$.

With this choice, the formula differs from our Rindler space discussion  only by a prefactor
(see equation (\ref{eqn:dellambdakU})), so we can copy the answer, continuing to neglect order one factors
\begin{equation}
   {1 \over T^2 V_1 V_2} \langle \Delta \lambda(V, k_\perp) \Delta \lambda(V', k_\perp') \rangle =
    G_N \delta^p(k_\perp + k_\perp') \int_0^\infty dk_V  e^{i k_V \Delta V}{ k_V \sin^2{k_\perp^2  U_* \over 4 k_V}\over k_\perp^4 }
\end{equation}
Consider the fluctuations of an operator smeared in $V$ by a small amount $\sigma_V$, with 
\begin{equation}
    \sigma_V \ll V
\end{equation}
Assuming Gaussian smearing for simplicity, this is given by
\begin{equation}
     {4 \over T^2 V^2} \langle \Delta \lambda_{\sigma_V}(V, k_\perp) \Delta \lambda_{\sigma_V}(V, k_\perp')\rangle =
    G_N {\delta^p(k_\perp + k_\perp')\over k_\perp^4} \int_0^\infty dk_V  e^{- k_V^2 \sigma_V^2}  k_V \sin^2{k_\perp^2  U_* \over 4 k_V}
\end{equation}
Rescaling the integration variable and rewriting the $\sin^2$ gives
\begin{equation}
    G_N {\delta^p(k_\perp + k_\perp')\over k_\perp^4 \sigma_V^2} \int_0^\infty dp  p e^{- p^2 }  \left(1 - \cos{k_\perp^2  U_* \sigma_V \over 2 p}\right)
\end{equation}
The integral depends on one dimensionless parameter 
\begin{equation}
b \equiv k_\perp^2 U_* \sigma_V
\end{equation}
When $b \gg 1$ the cosine term oscillates rapidly and can be ignored, so the integral gives 1. When $b \ll 1$, the analysis is more involved. We find
\begin{equation}
    \int_0^\infty dp p e^{-p^2}\left(1 - \cos{b \over  p}\right) = \begin{cases} 
    1 & b \gg 1 \\
   - b^2 \log b & b \ll 1 
   \end{cases}
\end{equation}
This gives the result for the perturbations
\begin{equation}
      \langle \Delta \lambda_{\sigma_V}(V, k_\perp)\Delta \lambda_{\sigma_V}(V, k_\perp')\rangle =
  \begin{cases} 
  G_N \delta^p(k_\perp + k_\perp'){\displaystyle T^2 V^2 \over k_\perp^4 \sigma_V^2} & b \gg 1 \\
 -  G_N \delta^p(k_\perp + k_\perp') T^2 V^2 U_*^2 \log (k_\perp^2 \sigma_V U_*) & b \ll 1
 \end{cases}
\end{equation}
The large $b$ limit is more reliable in that it is not dependent on the cutoff $U_*$, so the near-horizon approximation we have used here is self-consistent. For small $b$, the answer depends on the cutoff. Our result in this regime is more speculative. We can guess that the correct answer is given by fixing $U_*$ to be the place where the near-horizon approximation breaks down. Call the proper distance at which the near-horizon approximation breaks down $L_*$; this is related to $U_*$ by
\begin{equation}
    U_*^2 V^2 = L_*^4
\end{equation}
We would expect $L_*$ to be the Schwarzschild radius for Schwarzschild black holes; we discuss more general cases below. We then find
\begin{equation}
      \langle \Delta \lambda_{\sigma_V}(V, k_\perp)\Delta \lambda_{\sigma_V}(V, k_\perp')\rangle =
  \begin{cases} 
  G_N \delta^p(k_\perp + k_\perp'){\displaystyle T^2 V^2 \over k_\perp^4 \sigma_V^2} & b \gg 1 \\
 -  G_N \delta^p(k_\perp + k_\perp') T^2 L_*^4 \log (k_\perp^2 \sigma_V L_*^2/V) & b \ll 1
 \end{cases}
\end{equation}
We can rewrite our expression for the fluctuations in some different ways. First, we can relate the $V$ smearing $\sigma_V$ to a smearing in time. We have that
\begin{equation}
f'_0 dt = {dV \over V} - {dU \over U}
\end{equation}
while along constant $r$ we have 
\begin{equation}
{dU \over U} = -{dV \over V}
\end{equation}
Combining these gives
\begin{equation}
\Delta V = \Delta t V f'_0 \ \ \ \ \ \ \ \  ({\rm constant}\  r)
\end{equation}
where we have assumed that the smearing in $V$ is small,
\begin{equation}
{\sigma_V \over V} \ll 1 \ \ \ \ \leftrightarrow  \ \ \ \ {\sigma_t T}  \ll 1 \ \ \ \ \leftrightarrow \ \ \ \  {\deltatau^2 \over L_*^2} \ll 1
\end{equation}
Smearing over longer times would require going beyond our near-horizon approximation. This is particularly apparent in the rightmost inequality: for the near-horizon limit to be valid, the proper time for the experiment  must be much smaller than the characteristic length scale of the geometry $L_*$.

Using this to replace $\sigma_V $ by $\sigma_t$ we have, at small $b$,
\begin{equation} 
\label{flucktv}
\expval{(\Delta \lambda_h^{\sigma_t} (t, k_\perp))(\Delta \lambda_h^{\sigma_t} (t, k'_\perp))} = 
G_N \delta^{p}(k_\perp + k_\perp^\prime ){1 \over\sigma_t^2  k_\perp^4} 
\end{equation}
The  criterion $b \gg 1$ for the near-horizon approximation to be insensitive to the IR cutoff becomes
\begin{equation}
\left| L_*^2 T \sigma_t k_\perp^2 \right| \gg 1 \ \ \ \ \leftrightarrow \ \ \ \ k_\perp^2 \deltatau^2 \gg 1
\end{equation}

We can also translate from the continuous transverse momentum $k_\perp$ to discrete spherical harmonics. At large $\ell$, we have
\begin{equation}
k_\perp r_s \sim \ell \ \ \ \ \ \ \ \ \ \ \ \ \delta \lambda_h(V, \vec k_\perp) \sim r_s^p \delta \lambda_h(V, \vec \ell) \ \ \ \ \ \ \ \ \ \ \ \ \ \ \ \delta^p(k_\perp + k'_\perp) \sim {\delta_{\vec \ell, \vec \ell'} \over r_s^p}
\end{equation}
where  $\vec \ell$  denotes all of the angular momentum quantum numbers. The correlator becomes
\begin{equation}
\boxed {
\expval{\,\Delta \lambda_h^{\sigma_t} (t, \vec \ell)\, \Delta \lambda_h^{\sigma_t} (t, \vec \ell\,')\,} =
\begin{cases} \delta_{\vec \ell, \vec \ell'}
\displaystyle{G_N r_s^{4-p} \over\sigma_t^2 \ell^4} & \displaystyle{\ell \over r_s}  \deltatau \gg 1 \\
\delta_{\vec \ell, \vec \ell'}
{\displaystyle{G_N L_*^4 T^2 \over r_s^p}}  \log \left( \displaystyle{r_s \over \ell \deltatau} \right)&  \displaystyle{\ell \over r_s}  \deltatau \ll 1
\label{ellformula}
\end{cases} 
} 
\end{equation}

Note that, since we are at large $\ell$, we have only written the leading $\ell$ dependence. Recall that this formula is valid when the characteristic scales of the experiment are smaller than the curvature scale of the geometry,
\begin{equation}
\deltatau \ll L_* \ \ \ \ \ \ 
{\rm  and} \ \ \ \ \ \ 
\displaystyle{{r_s \over \ell}} \ll L_*
\end{equation}
Further, recall that the high-$\ell$ formula is insensitive to the IR cutoff $L_*$, while the low-$\ell$ formula depends on the cutoff, so the high-$\ell$ formula is more rigorous. 

\paragraph{Determining the IR cutoff $L_*$}
Looking at equation \eqref{nhmetric} for the near-horizon metric, we see that the Rindler approximation is valid if
\begin{equation}
\left| UV f'_0 r_s \right| \ll r_s^2 \ \ \ {\rm and} \ \ \ \ \left| f_0'' UV \right| \ll 1
\end{equation}
Note that the proper distance from the horizon is 
\begin{equation}
L^2 = - UV\
\end{equation}
so these requirements can be written simply in terms of the proper distance from the horizon where the approximation breaks down, $L_*$. The approximation is valid for 
\begin{equation}
L^2 \ll {r_s \over f'_0} \ \ \ {\rm and} \ \ \ \ L^2 \ll {1 \over |f_0''|}
\end{equation}
For black holes in asymptotically flat space, even if they are near-extremal, the Rindler approximation breaks down at 
\begin{equation}
L_*^2 \sim r_s^2
\end{equation}
while for large AdS black holes it is
\begin{equation}
L_*^2 \sim l_A^2
\end{equation}

\section{Calculating the Quantum Width}
In this section, we translate from the affine parameter fluctuations to the quantum width fluctuations.
Note that in the previous sections we have already defined a gauge-invariant observable and calculated (within certain approximations) its fluctuations. However, it is worthwhile to translate these fluctuations in the affine parameter to the horizon into a length scale, in order to gain intuition and relate our result to the literature. 

Roughly, the `quantum width' should be defined as the region near the horizon where it is uncertain whether it is inside or outside the horizon. Consider the following thought experiment: an ingoing radial null ray is sent towards the horizon, and after some affine parameter $\lambda_1$ a signal is sent back out towards infinity. The ingoing null ray does not know in advance what metric perturbations will be present; it is simply set to emit a signal at affine parameter $\lambda_1$.

Classically, the affine parameter to the horizon is a fixed value $\lambda_c$. If we choose $\lambda_1 < \lambda_c$, then the signal will escape the black hole; otherwise it will not.
However, we have now calculated the quantum fluctuations $\Delta \lambda_h$. Now if $\lambda_1$ is in the range
\begin{equation}
\lambda_c - \Delta \lambda_h < \lambda_1 < \lambda_c + \Delta \lambda_h
\end{equation}
the result is quite uncertain; the outgoing signal may or may not escape.

If we want our signal to have a good chance of escaping the black hole, it must be launched at an earlier affine parameter
\begin{equation}
\lambda_c - \Delta \lambda_h
\end{equation}
Typically, this will be an affine parameter $\Delta \lambda_h$ from the horizon. 

We have chosen the normalization of the affine parameter so that, in the background solution,
\begin{equation}
d \lambda = -d r
\end{equation}
near the horizon. So in this normalization, the quantum horizon reaches a coordinate distance
\begin{equation}
\Delta r = - \Delta \lambda_h
\end{equation}
from the event horizon. The formula for the size of these fluctuations is given in the previous sections.
We can translate this into a proper distance outside the horizon. The proper distance $L$ is given, in the near-horizon limit, by
\begin{equation}
L(\Delta r) = \int_{r_s}^{r_s+ \Delta r} {dr \over \sqrt{f}} \sim \sqrt{\Delta r \over T}
\end{equation}
Note that the quadratic relationship
\begin{equation}
L^2 \sim {\Delta r \over T}
\end{equation}
means that fluctuations are naturally related to $L^4$ rather than $L^2$. This is purely due to working with the proper distance.

To be clear:  we have defined the fluctuation in affine parameter carefully in a gauge invariant way. The conversion to proper distance is done for intuition and to compare to previous literature. From this point of view, the {\it definition} of the proper distance fluctuation is
\begin{equation}
    L^2 := {\Delta \lambda \over T}
\end{equation}

\subsection{Quantum Width as a function of angular momentum}
The quantum width for a given angular momentum mode $\ell$, measured with time resolution $\sigma_t$, is given by substituting the definition of $L$ into the formula for affine parameter fluctuations, \eqref{ellformula}, giving
\begin{equation}
\boxed{ \ \ \ \ \ \langle (L^2_{\ell, \sigma_t})^2 \rangle   = \begin{cases}
\displaystyle{G_N r_s^{4-p} \over T^2 \sigma_t^2 \ell^4} & \displaystyle{\ell \over r_s}  \deltatau \gg 1 \\
{\displaystyle{G_N L_*^4 \over r_s^p}}  \log \left( \displaystyle{r_s \over \ell \deltatau} \right)&  \displaystyle{\ell \over r_s}  \deltatau \ll 1
\end{cases} \ \ \ \ \ 
\label{eq-llfluc}}
\end{equation}
Note that the conversion to a proper length is only valid when the smearing is small compared to the distance from the horizon,
\begin{equation}
{\sigma_V \over V} \sim {\sigma_t T} \ll 1 \ \ \ \ \ \ \ {\rm Validity\ of\ proper\ distance}
\end{equation}
which is already the regime in which we have been calculating. Also note that two different time scales appear in the answer: the smearing in the initial Schwarzschild time of the experiment $\sigma_t$ and the proper time of the full experiment $\deltatau$. They are related by
\begin{equation}
    {\deltatau^2 \over L_*^2} = T \sigma_t 
\end{equation}
The fluctuations in affine parameter are well-defined when averaged over longer times, but the conversion to a proper distance is ambiguous.

%In addition, IR corrections to the formula are negligible when
%\begin{equation}
%{\omega_{\rm max} \over T} \ll {\ell^2 \over f_0'' r_s^2}
%\end{equation}
%Combining these constraints, our formula for the quantum width is valid when
%\begin{equation}
%1 \ll {\omega_{\rm max} \over T} \ll {\ell^2 \over f_0'' r_s^2}
%\end{equation}
%The quantum width can be rewritten 
%\begin{equation}
%L_\ell^4 =  {G_N r_s^{2-D} \over (f_0'')^2}\left( \omega_{\rm max}^2 (f_0'')^2 r_s^4 \over T^2 %%\ell^4 \right)
%\end{equation}
%The advantage of this way of writing it is that the factor in parentheses must be smaller than one for our approximation to be valid.

\subsection{Smearing in Angle}
So far we have worked in the angular momentum basis. Now we want to go back to real space on the sphere, and ask how the size of the fluctuations depends on the angular scale of the smearing. 

Smearing over an angular scale $\Delta \theta$ corresponds to summing over angular modes up to a maximum frequency \begin{equation}
\ell_{\rm max} \sim {1 \over \Delta \theta}
\end{equation}
We prefer to instead work in terms of the proper distance on the horizon over which we smear, 
\begin{equation}
    \sigma_\perp = r_s \Delta \ \ \ \ \ \rightarrow \ \ \ \ell_{\rm max} \sim {r_s \over \sigma_\perp}
    \end{equation}
The fluctuations of such an observable are given by
\begin{equation}
\langle (L^2_{\sigma_\perp, \deltatau})^2 \rangle  = \sum_{\vec \ell}^{\ell_{\rm max}} \langle (L_{\ell, \deltatau}^2)^2 \rangle 
\end{equation}
 Because the fluctuations of $L_{\ell, \deltatau}$ given in \eqref{eq-llfluc} take a different form in different regimes, evaluating the sum is somewhat tedious. For larger spatial smearing, $\sigma_\perp \gg \deltatau$, the `low-$\ell$' formula always dominates, and we find
 \begin{equation}
     {\langle (L^2_{\sigma_\perp, \deltatau})^2 \rangle\over L_*^4}  =  {G_N \over \sigma_\perp^p} \log{\sigma_\perp \over \deltatau} \ \ \ \ \ \ \ \ {\rm for} \ \ \ \sigma_\perp \gg \deltatau
 \end{equation}
In the regime of smaller spatial smearing $\sigma_\perp \ll \deltatau$, the result depends on the spatial dimension. When $D < 6$, corresponding to horizon dimension $p< 4$ the large $\ell$ part of the sum is convergent due to the $\ell^{-4}$ factor, so the result is independent of the spatial smearing
 \begin{equation}
     {\langle (L^2_{\sigma_\perp, \deltatau})^2 \rangle\over L_*^4 }  = {G_N \over \deltatau^p}  \ \ \ \ \ \ \quad \qquad {\rm for} \ \ \ \sigma_\perp \ll \deltatau \ \ \ {\rm and} \ \ \ p<4
 \end{equation}
In higher dimensions, the sum is dominated by the highest values of $\ell$ that are not suppressed by the angular smearing, giving 
 \begin{equation}
     {\langle (L^2_{\sigma_\perp, \deltatau})^2 \rangle \over L_*^4 } =  {G_N \over \sigma_\perp^{p-4}\deltatau^4}\ \ \ \ \ \ \ \qquad {\rm for} \ \ \ \sigma_\perp \ll \deltatau \ \ \ {\rm and} \ \ p>4
 \end{equation}
 In this regime we can re-express our result by holding fixed the smearing in Schwarzschild time $\sigma_t$. This leads to a result that is independent of the IR scale $L_*$ 
 \begin{equation}
       \langle (L^2_{\sigma_\perp, \sigma_t})^2 \rangle   = {G_N \over T^2 \sigma_t^2 \sigma_\perp^{p-4}} \ \ \ \ \  \qquad{\rm for} \ \ \ {\sigma^2_\perp \over T\sigma_t} \ll L^2_* \ \ \ {\rm and} \ \ p>4
 \end{equation}

\section{Discussion}

Before making some general concluding comments, let us first discuss the thermodynamic interpretation of our results,

\subsection{Thermodynamic Interpretation}

We would like to compare our results for the quantum width $L$ with the estimate (\ref{thermo-estimate}) based on black hole thermodynamics. Both expressions are proportional to $G_N$, but the logical reasoning behind their derivations are very different. In this paper, we performed a tree level perturbative calculation involving the two-point function of the graviton without making reference to any thermodynamic quantities. Nevertheless, we can write our results in way that is similar to the result (\ref{thermo-estimate}). 
For this purpose, let us introduce the entropy associated to a part of the horizon with transverse size $\sigma_\perp$
\begin{equation}
    S_{\sigma_\perp} = {\sigma_\perp^p \over 4 G_N}.
\end{equation}
Then the above equation can be rewritten
\begin{equation}
  % {\langle (\Delta U)^2 \rangle \over U_*^2} 
   {\bigl\langle\left(L^2_{\sigma_\perp, \deltatau}\right)^2\bigr\rangle \over L_*^4} 
   = {1 \over S_{\sigma_\perp}} \times    \begin{cases}
       {\displaystyle{ \left(  \sigma_\perp \over \deltatau  \right)}^p} & \sigma_\perp \ll \deltatau  \ \ {\rm and} \ \ p<4 \\
     {\displaystyle{ \left(  \sigma_\perp \over \deltatau  \right)}^4}&  \sigma_\perp \ll \deltatau \ \ {\rm and} \ \ p>4 \\
         \log {\sigma_\perp \over  \deltatau} & \sigma_\perp \gg \deltatau  
    \end{cases}
\end{equation}
Note that, aside from the universal $1/S$ term, the fluctuations are either suppressed by powers of the ratio of distance scales, or logarithmically enhanced, depending on the regime. 
Here we interpret the quantity $S_{\sigma_\perp}$, together with the other factors, as counting the entropy associated to the degrees of freedom whose fluctuations lead to the quantum width of the horizon.

%This has the form expected from thermodynamics, where fluctuations in any observable are controlled by the entropy,
%\begin{equation}
%    {(\Delta X)^2 \over X^2} \sim {1 \over S}
%\end{equation}

\subsection{Concluding comments}

In this paper, we computed the quantum width of the horizon for general spherically symmetric black holes as a function of the time and angular resolutions. We focused on the regimes in which the ratio of the transverse and longitudinal cut-offs is taken either to be very large or very small. In some regimes,  the result was found to depend on the infrared scale $L_*$, while in others the infrared dependence disappeared for space-time dimension greater than six. There are a number of future directions that would be very interesting to pursue:
\begin{itemize}
    \item {\bf Beyond the near-horizon approximation.} Our explicit calculations have been done in the approximation that the fluctuations are dominated by the near-horizon region and that this can be approximated as Rindler space. It would be very nice to go beyond this approximation, both to verify our results and to extend to regimes such as low frequency perturbations that are not accessible within our approximation.
    \item {\bf Infrared Issues and BMS.} We have thoroughly analyzed the infrared sensitivity of our observables within the near-horizon limit. However, it is important to understand infrared sensitivity beyond this limit. We expect that \begin{itemize}
        \item  For black holes in asymptotically Minkowski spacetime in 3+1 dimensions, the full analysis leads to logarithmic sensitivity to the IR cutoff. This could lead to a connection with the study of BMS symmetries and the celestial holography program \cite{Strominger:2017zoo, McLoughlin:2022ljp}.
        \item For black holes in asymptotically AdS spacetime, and for asymptotically flat black holes in $D>4$, we expect that the full observable is IR finite; the IR divergences found in our near-horizon analysis are a result of our approximation.
    \end{itemize} 
    \item{\bf More general horizons.} Our main goal was to calculate the quantum width of black hole horizons. But our methods apply equally well in other types of horizons, and have in fact been obtained for Rindler horizons. It would be interesting to extend this to even more general horizons, such as the cosmological horizon in de Sitter space, along the lines of \cite{Aalsma:2025bcg}. In flat space and in other maximally symmetric space-times, one can also consider finite causal diamonds that are conformally equivalent to the de Sitter static patch.  The boundaries of these causal diamonds behave in many ways as horizons and should therefore also have a quantum width that can be computed with methods similar to those in this work.  
    
    \item {\bf Near-Extremal Physics and the Schwarzian.} Recently, there has been much progress in understanding quantum gravity corrections to near-extremal black holes (see \cite{Turiaci:2024cad} for a review). What is the general relationship between our horizon fluctuations and the Schwarzian theory describing the low-energy fluctuations of near-extremal black holes?\footnote{We thank Suzanne Bintanja for bringing up this issue and for discussions.} A speculation is the horizon fluctuations we analyze are captured by the Schwarzian theory in a suitable low-energy, near-extremal limit. A concrete suggestion due to Steve Shenker is that our fluctuations are closely related to the `scramblon' physics described in \cite{Gu:2021xaj, Stanford:2023npy}. This limit is complementary to the regime where our near-horizon Rindler analysis is valid. 
    
    A related question is whether these fluctuations are enhanced in the near-extremal limit. This would require going beyond the Rindler approximation to treat the nearly $AdS_2$ geometry of near-extremal black holes.
    \item {\bf Connection to AdS/CFT and holographic hydrodynamics.} We have said nothing about the CFT interpretation of our observable. Since the fluctuations are given by an integral of gravitational perturbations, they are clearly related,  via the HKLL dictionary (see e.g. \cite{Kabat:2011rz}), to CFT integrals of the stress tensor fluctuations. Because we have a black hole in the bulk, we are interested in the stress tensor fluctuations around a thermal or microcanonical state. It remains to be seen whether our fluctuations have a simple CFT interpretation. A striking result is that the condition  for a given mode to be dominated by near-horizon physics is precisely related to the hydrodynamic sound mode \cite{Kovtun:2003wp} . Our criterion for insensitivity to the IR cutoff  is
\begin{equation}
{T \sigma_t k_\perp^2 \over |f_0''|} \gg 1
\end{equation}
To relate this to known AdS/CFT results, note that the time smearing $\sigma_t$ can be thought of as a frequency cutoff $\omega_{\rm max} \sim 1/\sigma_t$.  For large AdS black holes, this criterion is closely related to the hydrodynamic diffusion mode: if the diffusion mode has
\begin{equation}
\omega_{diff} = -i D k^2
\end{equation}
our criterion is
\begin{equation}
\omega_{\rm max} \ll D k^2
\end{equation}
There is a striking resemblence between the formulas. We expect that there is a sharp relationship between our horizon fluctuations and holographic hydrodynamics that is waiting to be uncovered.

\end{itemize}

% \subsection{Using transverse momentum to regulate the IR}
% It is interesting to also do this analysis in momentum space, where the IR regulator is more rigorously defined. Our basic formula from above is 
% \be
% \expval{\Delta \tilde u (0, v, k_\perp) \Delta \tilde u (0, v, k'_\perp)} = G_N {\delta^{D-2}(k_\perp + k'_\perp) \over k_\perp^4 } {1 \over \epsilon_v^2}
% \end{equation}
% We want to consider the horizon fluctuations at a given momentum scale. For this, the natural tool is the dimensionless power spectrum,
% \be
% \expval {(\Delta u(0, 0, 0))^2} = \int d^{D-2} k {G_N \over k_\perp^4 \epsilon_v^2} = \int_0^\infty {dk \over k}{G_N k^{D-6} \over \epsilon_v^2}
% \end{equation}
% which shows that the dimensionless power is
% \be
% \Delta^2(k_\perp) = {G_N \over \epsilon_v^2} k_\perp^{D-6}
% \end{equation}

% We use the same logic as before: the smearing in $v$ is constrained by

\subsection*{Acknowledgements}
It is a pleasure to thank Lars Aalsma,  Suzanne Bintanja, Beatrice Bonga, Raphael Bousso, Latham Boyle, Bruno Bucciotti, Dan Carney, Yanbei Chen, Laurent Friedel, Steve Giddings, Temple He, Arthur Hebecker, Diego Hofman, J\"org J\"ackel, Manthos Karydas, Cindy Keeler, Renate Loll, Mark Mezei, Richard Myers, Rob Myers, Maulik Parikh, Don Marolf, Prahar Mitra, Upamanyu Moitra, Rob Myers, Andrea Puhm, Steve Shenker, Eva Silverstein, Lenny Susskind, Manus Visser, and Kathryn Zurek. Apologies to anyone we have forgotten to list. This work was partially supported by Heising-Simons Foundation ‘Observational Signatures of Quantum Gravity’ QuRIOS collaboration grant. 

\appendix

\section{Coordinates for static black holes}
In Schwarzschild coordinates, the metric is
\begin{equation}
ds^2 = - f(r) dt^2 + {dr^2 \over f(r)} + r^2 d\Omega^2
\end{equation}
Define the Eddington-Finkelstein coordinate $v$ by
\begin{equation}
dv = dt + {dr \over f(r)}
\end{equation}
In Eddington-Finkelstein coordinates, the metric becomes
\begin{equation}
ds^2 = -f(r) dv^2 + 2 dr dv + r^2 d\Omega^2
\end{equation}
These coordinates work well at the future horizon, but $v \to -\infty$ at the past horizon. To deal with this, define the Kruskal coordinate $V$ by
\begin{equation}
V = r_s e^{v f'_0/ 2 }
\end{equation}
where $f'_0 \equiv f'(r_s)$ is the Hawking temperature (times $4\pi$). 
Define $U$ by
\begin{equation}
-UV = {2(r_0-r_s) \over f'_0} \exp \left(f'_0 \int_{r_0}^r {dr' \over f(r')} \right)
\label{eq-rdef}
\end{equation}
where $r_0$ is an arbitrary location very close to the black hole horizon.

With these choices, we have
\begin{equation}
{dV \over V} = {f'_0 \over 2} dv \ \ \ \ \text{and} \ \ \ \ {dU \over U} + {dV \over V} = {f_0' \over f(r)} dr
\end{equation}
As a result, the metric takes the form
\begin{equation}
ds^2 = {4 f(r) \over (f'_0)^2 UV} dU dV + r^2 d\Omega^2
\end{equation}
with $r$ defined by \eqref{eq-rdef}.

In the near horizon limit, we have 
\begin{equation}
-UV \approx {2 \over f'_0}(r- r_s) + \dots
\end{equation}
and after careful near-horizon expansions, the metric becomes
\begin{equation}
\label{nhmetric}
ds^2 \approx (- 2 + f''_0 UV+ \dots) dU dV + (r_s^2 - f'_0 r_s UV + \dots ) d\Omega^2
\end{equation}
Note that the two correction terms can be very different for near-extremal RN black holes. For Schwarzschild, $f_0' = 1/r_s$ and $f_0'' = -2/r_s^2$ so
\begin{equation}
ds^2 \approx (- 2 - 2{ UV \over r_s^2}+ \dots) dU dV + (r_s^2 -  UV + \dots ) d\Omega^2 \ \ \ \ \ \ \text{(Schwarzschild)}
\end{equation}
For the purpose of transforming tensors, it is useful to collect the relation between the coordinates. Between Kruskal and Eddington Finkelstein we have
\begin{eqnarray}
\label{k2ef}
dv &=& {2 \over f'_0} {dV \over V} \\
{f'_0 \over f} dr &=& {dU \over U} + {dV \over V}
\end{eqnarray}
or equivalently
\begin{eqnarray}
\label{ef2k}
{dV \over V} &=& {f_0' \over 2} dv \\
{dU \over U} &=& {f_0' \over f} dr - {f_0' \over 2} dv
\end{eqnarray}
Relating Schwarzschild to Kruskal gives
\begin{eqnarray}
f'_0 dt &=& {dV \over V} - {dU \over U} \\
    {f'_0 \over f} dr &=& {dU \over U} + {dV \over V}
\end{eqnarray}
or, equivalently
\begin{eqnarray}
    2 {dV \over V} = f'_0({dr \over f} + dt) \\
    2 {dU \over U} = f'_0({dr \over f} - dt)
\end{eqnarray}

\section{$\mathscr{I}^+$ normalized observable}
\label{app:scri+}

Here we include some computations
of the modified observable defined in
equation (\ref{eqn:scri+norm}).   In this case
the two-point function evaluates to
\begin{equation}
    \langle \Delta \lambda (k_V, k_\perp) \Delta \lambda(k_V', k_\perp') \rangle = G_N \delta^p(k_\perp + k') \delta(k_V + k_V') \int_0^\infty  dk_U \left|{1 - e^{i k_U U_*} \over i k_U} + U_* e^{i k_U U_*} \right|^2 \delta (2 k_U k_V - k_\perp^2)
\end{equation}
The integrand contains a term that grows quadratically with $U_*$ at large $U_*$. So the large cutoff limit of this experiment does not 
agree with the observable analyzed in 
section \ref{sec:rindler}. This is somewhat surprising, because if one assumes that fluctuations go to zero at infinity, one might expect that it will not matter which choice of normalization we use in the limit.

If we look at the momentum space correlator, it has a simple form in two limits, 
\begin{equation}
    \langle \Delta \lambda (k_V, k_\perp) \Delta \lambda(k_V', k_\perp') \rangle \approx G_N \delta^p(k_\perp + k') \delta(k_V + k_V') \Theta(k_V) \times \begin{cases}
        {U_*^4 k_\perp^4 \over k_V^3} & {k_\perp^2 U_* \over k_V} \ll 1 \\
        {U_*^2 \over k_V} & {k_\perp^2 U_* \over k_V} \gg 1
    \end{cases}
    \end{equation}
    This equation differs substantially from the corresponding equation in the other normalization. 
    
    However, when we calculate the fluctuations of a smeared operator, the form of the answer is more similar. Here, to get a finite result, we need to tame potential divergences that occur at $k_V \to 0$, corresponding because of the on-shell condition to $k_U \to \infty$. We deal with these by smearing slightly in $U$. Assuming the smearing in $U$ is smaller than other scales in the problem, we have
    \begin{equation}
         \langle (\Delta \lambda_{\sigma_V, \sigma_\perp} (U_*) )^2 \rangle =   \begin{cases}
         {\displaystyle G_N U_*^2 \over \sigma_\perp^p} \log{\sigma_U \sigma_V \over \sigma_\perp^2} & \sigma_\perp^2 \ll U_* \sigma_V \\
         {\displaystyle G_N U_*^2 \over \sigma_\perp^p} \log {U_* \over \sigma_U} & \sigma_\perp^2 \gg U_* \sigma_V 
         \end{cases}
    \end{equation}
 %   If we write this in a way that is reminiscent of thermodynamics, as above, we find simply
 %   \begin{equation}
 %        {\langle (\Delta U_{\sigma_V, \sigma_\perp}  )^2 \rangle \over U_*^2}  \sim {1 \over S_{\sigma_\perp}} \times  \log (\dots ) 
 %        \end{equation}
%so the fluctuations of this observable are also given by thermodynamics in a simple way, up to logarithms.

It would be nice to have a clearer physical understanding of the relationship between our two different finite cutoff experiments in the limit that the cutoff is taken away,  $U_* \to \infty$.

\bibliography{bh-flucs-ref.bib}

@article{Marolf2003,
    author = "Marolf, Donald",
    editor = "Trampetic, Josip and Wess, Julius",
    title = "{On the quantum width of a black hole horizon}",
    eprint = "hep-th/0312059",
    archivePrefix = "arXiv",
    doi = "10.1007/3-540-26798-0_9",
    journal = "Springer Proc. Phys.",
    volume = "98",
    pages = "99--112",
    year = "2005"
}

@article{Kovtun:2003wp,
    author = "Kovtun, Pavel and Son, Dam T. and Starinets, Andrei O.",
    title = "{Holography and hydrodynamics: Diffusion on stretched horizons}",
    eprint = "hep-th/0309213",
    archivePrefix = "arXiv",
    reportNumber = "UW-PT-03-21, INT-PUB-03-17",
    doi = "10.1088/1126-6708/2003/10/064",
    journal = "JHEP",
    volume = "10",
    pages = "064",
    year = "2003"
}

@article{Turiaci:2024cad,
    author = "Turiaci, Gustavo Joaquin",
    title = "{Les Houches lectures on two-dimensional gravity and holography}",
    eprint = "2412.09537",
    archivePrefix = "arXiv",
    primaryClass = "hep-th",
    doi = "10.21468/SciPostPhysLectNotes.113",
    journal = "SciPost Phys. Lect. Notes",
    volume = "113",
    pages = "1",
    year = "2026"
}

@article{Kabat:2011rz,
    author = "Kabat, Daniel and Lifschytz, Gilad and Lowe, David A.",
    title = "{Constructing local bulk observables in interacting AdS/CFT}",
    eprint = "1102.2910",
    archivePrefix = "arXiv",
    primaryClass = "hep-th",
    doi = "10.1103/PhysRevD.83.106009",
    journal = "Phys. Rev. D",
    volume = "83",
    pages = "106009",
    year = "2011"
}

@article{Gu:2021xaj,
    author = "Gu, Yingfei and Kitaev, Alexei and Zhang, Pengfei",
    title = "{A two-way approach to out-of-time-order correlators}",
    eprint = "2111.12007",
    archivePrefix = "arXiv",
    primaryClass = "hep-th",
    doi = "10.1007/JHEP03(2022)133",
    journal = "JHEP",
    volume = "03",
    pages = "133",
    year = "2022"
}

@article{Stanford:2023npy,
    author = "Stanford, Douglas and Vardhan, Shreya and Yao, Shunyu",
    title = "{Scramblon loops}",
    eprint = "2311.12121",
    archivePrefix = "arXiv",
    primaryClass = "hep-th",
    doi = "10.1007/JHEP10(2024)073",
    journal = "JHEP",
    volume = "10",
    pages = "073",
    year = "2024"
}

@book{Strominger:2017zoo,
    author = "Strominger, Andrew",
    title = "{Lectures on the Infrared Structure of Gravity and Gauge Theory}",
    eprint = "1703.05448",
    archivePrefix = "arXiv",
    primaryClass = "hep-th",
    isbn = "978-0-691-17973-5",
    publisher = "Princeton University Press",
    year = "2018"
}

@article{McLoughlin:2022ljp,
    author = "McLoughlin, Tristan and Puhm, Andrea and Raclariu, Ana-Maria",
    title = "{The SAGEX review on scattering amplitudes chapter 11: soft theorems and celestial amplitudes}",
    eprint = "2203.13022",
    archivePrefix = "arXiv",
    primaryClass = "hep-th",
    reportNumber = "SAGEX-22-12, CPHT-RR016.032022, HU-EP-22/13, TCDMATH 22-02",
    doi = "10.1088/1751-8121/ac9a40",
    journal = "J. Phys. A",
    volume = "55",
    number = "44",
    pages = "443012",
    year = "2022"
}

@article{Parikh:2024zmu,
    author = "Parikh, Maulik and Pereira, Jude",
    title = "{Quantum uncertainty in the area of a black hole}",
    eprint = "2412.21160",
    archivePrefix = "arXiv",
    primaryClass = "hep-th",
    doi = "10.1007/JHEP09(2025)137",
    journal = "JHEP",
    volume = "09",
    pages = "137",
    year = "2025"
}

@misc{btz, 
    author = {Freivogel, Ben and Moitra, Upamanyu}, 
    year   = {(2026)},
    howpublished = {\emph{To appear}}
    }

@article{Hu:2006dd,
    author = "Hu, B. L. and Roura, Albert",
    title = "{Fluctuations of an evaporating black hole from back reaction of its Hawking radiation: Questioning a premise in earlier work}",
    eprint = "gr-qc/0601088",
    archivePrefix = "arXiv",
    doi = "10.1007/s10773-007-9338-x",
    journal = "Int. J. Theor. Phys.",
    volume = "46",
    pages = "2204--2217",
    year = "2007"
}

@inproceedings{Hu:2006jc,
    author = "Hu, B. L. and Roura, Albert",
    title = "{Black hole fluctuations and dynamics from back-reaction of Hawking radiation: Current work and further studies based on stochastic gravity}",
    booktitle = "{7th Asia-Pacific International Conference on Gravitation and Astrophysics (ICGA7 2005)}",
    eprint = "gr-qc/0610066",
    archivePrefix = "arXiv",
    doi = "10.1142/9789812772923_0029",
    pages = "236--250",
    month = "10",
    year = "2006"
}

@article{Ciambelli:2026pwi,
    author = "Ciambelli, Luca and He, Temple and Klinger, Marc S. and Zurek, Kathryn M.",
    title = "{Mapping the Infrared Phase Space of Gravity to Finite Subregions}",
    eprint = "2606.12515",
    archivePrefix = "arXiv",
    primaryClass = "hep-th",
    reportNumber = "CALT-TH 2026-023",
    month = "6",
    year = "2026"
}

@article{Ciambelli:2025fbo,
    author = "Ciambelli, Luca and He, Temple and Zurek, Kathryn M.",
    title = "{From Asymptotically Flat Gravity to Finite Causal Diamonds}",
    eprint = "2512.09018",
    archivePrefix = "arXiv",
    primaryClass = "hep-th",
    reportNumber = "CALT-TH 2025-039",
    doi = "10.1103/lbm1-vkks",
    journal = "Phys. Rev. Lett.",
    volume = "136",
    number = "19",
    pages = "191501",
    year = "2026"
}

@article{He:2025hag,
    author = "He, Temple and Mitra, Prahar and Zurek, Kathryn M.",
    title = "{Effective Density Matrix for Vacua in Asymptotically Flat Gravity}",
    eprint = "2509.13401",
    archivePrefix = "arXiv",
    primaryClass = "hep-th",
    reportNumber = "CALT-TH 2025-030",
    doi = "10.1103/n48t-3dr1",
    journal = "Phys. Rev. Lett.",
    volume = "136",
    number = "21",
    pages = "211501",
    year = "2026"
}

@article{Fransen:2025npa,
    author = "Fransen, Kwinten and He, Temple and Zurek, Kathryn M.",
    title = "{Thermodynamics of a spherically symmetric causal diamond in Minkowski spacetime}",
    eprint = "2507.22977",
    archivePrefix = "arXiv",
    primaryClass = "hep-th",
    reportNumber = "CALT-TH 2025-026",
    doi = "10.1007/JHEP12(2025)125",
    journal = "JHEP",
    volume = "12",
    pages = "125",
    year = "2025"
}

@article{Zhang:2023mkf,
    author = "Zhang, Yiwen and Zurek, Kathryn M.",
    title = "{Stochastic description of near-horizon fluctuations in Rindler-AdS}",
    eprint = "2304.12349",
    archivePrefix = "arXiv",
    primaryClass = "hep-th",
    doi = "10.1103/PhysRevD.108.066002",
    journal = "Phys. Rev. D",
    volume = "108",
    number = "6",
    pages = "066002",
    year = "2023"
}

@article{Gukov:2022oed,
    author = "Gukov, Sergei and Lee, Vincent S. H. and Zurek, Kathryn M.",
    title = "{Near-horizon quantum dynamics of 4D Einstein gravity from 2D Jackiw-Teitelboim gravity}",
    eprint = "2205.02233",
    archivePrefix = "arXiv",
    primaryClass = "hep-th",
    reportNumber = "CALT-TH-2022-016",
    doi = "10.1103/PhysRevD.107.016004",
    journal = "Phys. Rev. D",
    volume = "107",
    number = "1",
    pages = "016004",
    year = "2023"
}

@article{Banks:2021jwj,
    author = "Banks, Thomas and Zurek, Kathryn M.",
    title = "{Conformal description of near-horizon vacuum states}",
    eprint = "2108.04806",
    archivePrefix = "arXiv",
    primaryClass = "hep-th",
    doi = "10.1103/PhysRevD.104.126026",
    journal = "Phys. Rev. D",
    volume = "104",
    number = "12",
    pages = "126026",
    year = "2021"
}

@article{Zurek:2020ukz,
    author = "Zurek, Kathryn M.",
    title = "{On vacuum fluctuations in quantum gravity and interferometer arm fluctuations}",
    eprint = "2012.05870",
    archivePrefix = "arXiv",
    primaryClass = "hep-th",
    doi = "10.1016/j.physletb.2022.136910",
    journal = "Phys. Lett. B",
    volume = "826",
    pages = "136910",
    year = "2022"
}

@article{Verlinde:2019ade,
    author = "Verlinde, Erik and Zurek, Kathryn M.",
    title = "{Spacetime Fluctuations in AdS/CFT}",
    eprint = "1911.02018",
    archivePrefix = "arXiv",
    primaryClass = "hep-th",
    doi = "10.1007/JHEP04(2020)209",
    journal = "JHEP",
    volume = "04",
    pages = "209",
    year = "2020"
}

@article{Verlinde:2019xfb,
    author = "Verlinde, Erik P. and Zurek, Kathryn M.",
    title = "{Observational signatures of quantum gravity in interferometers}",
    eprint = "1902.08207",
    archivePrefix = "arXiv",
    primaryClass = "gr-qc",
    doi = "10.1016/j.physletb.2021.136663",
    journal = "Phys. Lett. B",
    volume = "822",
    pages = "136663",
    year = "2021"
}

@article{Freidel:2026hed,
    author = "Freidel, Laurent and Oberfrank, Robin",
    title = "{Geometric noise spectrum in interferometers}",
    eprint = "2601.17849",
    archivePrefix = "arXiv",
    primaryClass = "hep-th",
    month = "1",
    year = "2026"
}

@article{Aalsma:2025bcg,
    author = "Aalsma, Lars and Bak, Sang-Eon",
    title = "{Modular fluctuations in cosmology}",
    eprint = "2503.04886",
    archivePrefix = "arXiv",
    primaryClass = "hep-th",
    doi = "10.1103/cpyr-w532",
    journal = "Phys. Rev. D",
    volume = "112",
    number = "2",
    pages = "026017",
    year = "2025"
}

@article{Carney:2024wnp,
    author = "Carney, Daniel and Karydas, Manthos and Sivaramakrishnan, Allic",
    title = "{Response of interferometers to the vacuum of quantum gravity}",
    eprint = "2409.03894",
    archivePrefix = "arXiv",
    primaryClass = "hep-th",
    reportNumber = "Phys. Rev. D 113, 106002 (2026)",
    doi = "10.1103/j5kj-zdky",
    journal = "Phys. Rev. D",
    volume = "113",
    number = "10",
    pages = "106002",
    year = "2026"
}

@article{Bak:2023wwo,
    author = "Bak, Sang-Eon and Parikh, Maulik and Sarkar, Sudipta and Setti, Francesco",
    title = "{Quantum-gravitational null Raychaudhuri equation}",
    eprint = "2312.17214",
    archivePrefix = "arXiv",
    primaryClass = "gr-qc",
    doi = "10.1007/JHEP07(2024)214",
    journal = "JHEP",
    volume = "07",
    pages = "214",
    year = "2024"
}

@article{Bak:2022oyn,
    author = "Bak, Sang-Eon and Parikh, Maulik and Sarkar, Sudipta and Setti, Francesco",
    title = "{Quantum gravity fluctuations in the timelike Raychaudhuri equation}",
    eprint = "2212.14010",
    archivePrefix = "arXiv",
    primaryClass = "gr-qc",
    doi = "10.1007/JHEP05(2023)125",
    journal = "JHEP",
    volume = "05",
    pages = "125",
    year = "2023"
}

@article{Bonga:2013uha,
    author = "Bonga, B{\'e}atrice and Khavkine, Igor",
    title = "{Quantum astrometric observables II: time delay in linearized quantum gravity}",
    eprint = "1307.0256",
    archivePrefix = "arXiv",
    primaryClass = "gr-qc",
    reportNumber = "ITP-UU-13-16, SPIN-13-11",
    doi = "10.1103/PhysRevD.89.024039",
    journal = "Phys. Rev. D",
    volume = "89",
    number = "2",
    pages = "024039",
    year = "2014"
}

@article{Banks:2023wua,
    author = "Banks, T. and Fischler, W.",
    title = "{Fluctuations and Correlations in Causal Diamonds}",
    eprint = "2311.18049",
    archivePrefix = "arXiv",
    primaryClass = "hep-th",
    month = "11",
    year = "2023"
}

@article{Banks:2025erx,
    author = "Banks, Tom",
    title = "{The hydrodynamic approach to quantum gravity}",
    eprint = "2505.15941",
    archivePrefix = "arXiv",
    primaryClass = "hep-th",
    reportNumber = "RUNHETC-2025-13",
    doi = "10.1142/S0218271825440201",
    journal = "Int. J. Mod. Phys. D",
    volume = "34",
    number = "16",
    pages = "2544020",
    year = "2025"
}
\bibliographystyle{JHEP-thesis}

\end{document}